# Cratered Lorentzian response of driven microwave superconducting nanowire-bridged resonators: oscillatory and magnetic-field induced stochastic states

Matthew W. Brenner[1], Sarang Gopalakrishnan[1,2,3], Jaseung Ku[1], Timothy J. McArdle[1], James N. Eckstein[1], Nayana Shah[4], Paul M. Goldbart[1,2], and Alexey Bezryadin[1]

[1] Department of Physics, University of Illinois at Urbana-Champaign, Urbana, Illinois 61801, U.S.A.
[2] Institute for Condensed Matter Theory, University of Illinois at Urbana-Champaign, Urbana, Illinois 61801, U.S.A.
[3] Kavli Institute for Theoretical Physics, University of California, Santa Barbara, California 93106, U.S.A.
[4] Department of Physics, University of Cincinnati, P.O. Box 210011, Cincinnati, Ohio 45221, U.S.A.

**Abstract** — Microwave Fabry-Perot resonators containing nonlinear mesoscopic elements (such as superconducting nanowires) can be used to explore many-body circuit QED. Here, we report on observations of a superconductor-normal pulsing regime in microwave (GHz) coplanar waveguide resonators consisting of superconducting MoGe films interrupted by a gap that is bridged by one or more suspended superconducting nanowires. This regime, which involve MHz-frequency oscillations in the amplitude of the supercurrent in the resonator, are achieved when the *steady-state* amplitude of the current in the driven resonator exceeds the critical current of the nanowires. Thus we are able to determine the temperature dependence of the critical current, which agrees well with the corresponding Bardeen formula. The pulsing regime manifests itself as an apparent "crater" on top of the fundamental Lorentzian peak of the resonator. Once the pulsing regime is achieved at a fixed drive power, however, it remains stable for a range of drive frequencies corresponding to subcritical steady state currents in the resonator. We develop a phenomenological model of resonator-nanowire systems, from which we are able to obtain a quantitative description of the amplitude oscillations and also, *inter alia*, to investigate thermal relaxation processes in superconducting nanowires. For the case of resonators comprising *two* parallel nanowires and subject to an external magnetic field, we find field-driven oscillations of the onset power for the amplitude oscillations, as well as the occurrence (for values of the magnetic field that strongly frustrate the nanowires) of a distinct steady state in which the pulsing is replaced by stochastic amplitude-fluctuations. We conclude by giving a brief discussion of how circuit-QED-based systems have the potential to facilitate nondestructive measurements of the current-phase relationship of superconducting nanowires and, hence, of the rate at which quantum phase-slips take place in superconducting nanowires.

# 1. Introduction

A variety of recent advances in quantum computing and related fields have involved circuit-QED systems [1, 2], which consist of mesoscopic elements (e.g., "artificial atoms") that are strongly coupled to microwave resonators. Circuit-QED has the advantage over *cavity* QED [3], which involves atoms coupled to an optical cavity, in that the strong-coupling regime, in which the resonance frequency shifts appreciably when a single photon is added to the cavity, is easier to attain [1]. The condition for cavity QED effects to become significant is that the relevant excitation energy scales of the "atomic" system should be similar to those of a single cavity photon. To date, most of the mesoscopic elements employed in circuit-QED settings have been "artificial atoms" (e.g., Cooper-pair boxes [4]), capacitively coupled to the resonator; in addition, a Josephson junction has, in certain experiments, been integrated into the resonator itself in order to facilitate the "weak" measurement [5] of the internal states of the artificial atoms [6, 7, 8]. Both artificial atoms and Josephson junctions are essentially zero-dimensional quantum systems; the present work is motivated by the possibility of extending the circuit-QED paradigm to spatially extended systems, which have more complex internal structure and thus richer excitation spectra. In particular, we focus on superconducting nanowires, which are believed to exhibit many-body phenomena such as Little's phase slips [9], which can occur either by thermal activation or by quantum tunneling [10, 11, 12, 13, 14, 15]. It was recently shown that superconducting nanowires act as nonlinear inductive elements, so they have been proposed as building blocks for qubits [16]. Circuit-QED systems involving superconducting nanowires thus raise the possibility of bringing the physics of *many-body cavity QED*—which involves, e.g., Bose-Einstein condensates (BECs) coupled to optical cavities [17, 18, 19, 20]—to the solid-state setting. It is known that, in the context of atomic and optical physics, the cavity QED element offers new routes for probing the quantum dynamics of the BEC [21], as well as generating coupled matter-light phases that cannot be achieved with the BEC alone [19, 20]. Specifically, resonator-induced collective behavior, which has been studied in the cavity QED case and which we focus on in the present work, has no direct analog, to date, in circuit-QED.

Further motivation for the present work comes from the following points. First, the strategies used to date to probe quantum phase slips in superconducting nanowires are complicated by the need to measure very small resistances. Other experiments have bypassed this difficulty by studying the phase-slip induced formation of a Joule-heated quasi-normal state [13, 14, 15]. The new avenues opened up by circuit-QED-based experiments should enable the complications of measuring small resistances to be by-passed and compliment the studies done by measuring switching rates into the Joule-heated quasi-normal state. In addition, working with superconducting nanowires, rather than oxide-based superconducting tunnel junctions, should ameliorate complications due to trapped charges in the oxide material.

Thin-film superconducting resonators have been extensively studied [22, 23, 24] including a situation that incorporates a Josephson junction as a nonlinear inductive element into the resonator [25]. In the present work we study systems involving a microwave coplanar waveguide resonator having either one or two superconducting nanowires integrated into it [16] (see Fig. 1). The nanowires, as well as the central conductor of the resonator, are made of MoGe, and are fabricated by molecular templating, as discussed in Refs. [12, 26]. The nanowires are suspended over a trench, rather than resting on a substrate. The nanowire-resonator systems that we have fabricated can, in principle, be cooled to the cavity QED regime, as the following heuristic argument shows. The energy cost of a phase slip vanishes as the current through the nanowire approaches its critical value. The r.m.s. amplitude of the antinodal current corresponding to a single photon is given by the equipartition theorem as $\langle LI^2/2 \rangle = \hbar\omega_0/2$: here, $\omega_0$ is the resonator's natural frequency ($\omega_0/2\pi$ = 10 GHz) and $L$ = 1 nH is the order of magnitude of a typical circuit inductance, corresponding to an impedance of $Z_0$ = 50 Ω) [16]. For these values, $I_{\text{r.m.s.}} \approx \omega_0\sqrt{\hbar/Z_0} \cong 100$ nA , which is of the same order of magnitude as the critical current of a nanowire: e.g., a critical current of ~200 nA was previously reported [27, 28]. Thus, the *quantum* regime is, in principle, achievable. This regime would, e.g., feature coherent superpositions of states having distinct currents and cavity-photon numbers.

Here, we present measurements on a microwave coplanar waveguide resonator containing one or two superconducting nanowires at various temperatures and photon populations. These measurements, though not in the cavity QED regime *per se*, nevertheless reveal several puzzling features that are related to the physics of superconducting nanowires, and that ought to be understood and accounted for before the *quantum* dynamics of the composite system is addressed. These phenomena are related to the existence, for a strongly driven resonator, of a nonequilibrium, time-domain pulsing regime (which, in the frequency domain, we term the "crater") in which the nanowire is found to switch, periodically, between its normal and superconducting states. (resonators integrated with micro-bridges have been studied in [29]) We develop a simple, quantitative model of this pulsing regime, which captures all of its salient features; the success of our model poses constraints on how far neglected features, such as supercurrent dissipation can influence the behavior of a nanowire in a resonator. Furthermore, we find that for resonators that have *two* nanowires embedded in them, the threshold for entering the pulsing state depends periodically on the magnetic field perpendicularly applied, with a period consistent with the predictions and observations given in Refs. [30, 31]. Finally, we report on two tantalizing phenomena. First, the unexpected rise of the crater floor as its width approaches the maximum value, which occurs at the magnetic field such a phase difference of π/2 is induced between the ends of the wire (i.e. at the maximally frustrated state of the device). This frustrated state is associated with the lowest critical-power value where the energy of the system is the same for the states having a vortex number *n* and *n* + 1. Second, the appearance, at low temperatures and drive powers, of jumps in the resonance frequency, which are suggestive of a multivalued current-phase relationship in the nanowire. We will present a qualitative discussion of these phenomena.

This paper is organized as follows: In Section 2 we discuss experimental details involving the fabrication of the resonator and the design of the circuit. In the next two sections we discuss the case of a resonator having a single nanowire: in particular, in Section 3 we summarize the main features of the transmission characteristics for a resonator featuring a single wire; and in Section 4 we develop a phenomenological model that fits the transmission characteristics, and we discuss the extent to which the transmission characteristics contain information about the internal structure of the nanowire. In Section 5 we turn to the case of resonators that incorporate *two* nanowires. These devices exhibit a range of unanticipated effects, as the (perpendicular) magnetic field is varied. Finally, in Section 6 we present our conclusions and discuss the outlook for future work involving resonators having embedded nanowires.

## 2. Sample fabrication and experimental design

The nanowire-resonator samples were fabricated using the molecular templating technique [12, 26]. The nanowire is integrated into a superconducting coplanar waveguide (CPW) resonator using optical lithography. To make the nanowire, single-walled carbon nanotubes were deposited on a Si-SiO$_2$-SiN substrate, which contained a 100 nm wide trench across the center of the chip. The trench is produced through a process involving e-beam lithography, reactive ion etching, and wet etching in HF (to produce an undercut). The trench was aligned with the center of the resonator, in order to create a gap in the resonator's center conductor [see Fig. 1(a)]. Then a thin film (here 10 or 25 nm) of Mo$_{0.76}$Ge$_{0.24}$ (from Super Conductor Materials Inc.) was deposited across the surface of the sample using an AJA DC magnetron sputtering system (ATC 2000 from AJA International Inc.). The nanotubes that cross the trench became substrates for superconducting MoGe nanowires, as a result of the metal sputtering. Following Boaknin *et al.* [25], the resonator was patterned by photolithography, and the photo-mask was positioned so that just one or two nanowire(s) connect the two halves of the center conductor which, as mentioned above, is interrupted by the trench (see Fig. 1).

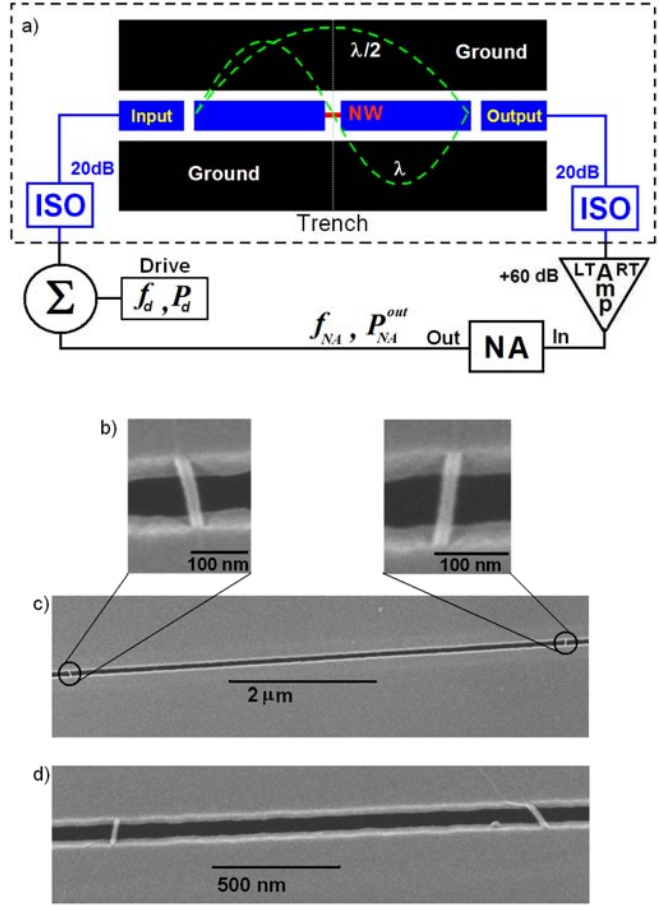

*Fig. 1.* (a) Resonator-nanowire schematic and measurement setup. The microwave signal, composed of one or two sinusoidal waves [added through a power combiner (denoted ∑)], is directed to the input of the Fabry-Perot resonator through a total of ~30 dB of isolators and attenuators, that are maintained at a cryogenic temperature. The input and output of the resonator are capacitively coupled to the center conductor (blue) through coupling capacitors of ~45 fF, formed by a 3 μm gap in the MoGe film. The center conductor is either 10 or 25 nm thick and 10 μm wide. It is interrupted, halfway along its length, by a trench, and connected through the nanowire(s). The output signal of the resonator travels through ~20 dB of isolators and attenuators (also at a cryogenic temperature) and then ~60 dB of amplifiers, including one at a cryogenic temperature, before arriving at the input port of the network analyzer (NA), where the transmitted power is measured. The spatial profiles of the supercurrents corresponding to the $\lambda/2$ and $\lambda$ modes are shown as dashed lines. (b) Examples of the individual, 25 nm wide, MoGe nanowires from double-nanowire sample S5. The trench, over which the wires are suspended, appears black. (c) Double-nanowire sample S5, showing the pair of nanowires, which appear geometrically similar. (d) Double-nanowire sample, S6, showing the pair of nanowires, which appear geometrically somewhat different.

This fabrication technique results in high-quality nanowires, which seamlessly connect the two halves of the resonator. The center conductor of the resonator is either 10 or 25 nm thick and ~10 μm wide, and the gap between the ground plane and center conductor is ~5 μm. A Fabry-Perot resonator is formed by gaps of ~3 μm between the center conductor and the input and output ports of the resonator. These gaps form two capacitors having capacitances of about 45 fF each, which act as two semi-transparent mirrors to impose a rigid boundary condition such that the supercurrent through these gaps is exactly zero. The total length of the center conductor between the two coupling gaps is 10 mm and the expected fundamental resonant frequency was ~10 GHz; however, the measured resonant frequency at low temperature was ~4 GHz, due to the kinetic inductance contributed by the MoGe film. All samples were designed to be overcoupled to have quality factors that are dominated by external dissipation from the energy leakage thorough the capacitive mirrors to the input and output ports, rather than by internal dissipation in the resonators.

Figure 1(b) shows scanning electron micrograph (SEM) images of typical superconducting MoGe nanowires (here having lengths and widths of ~105 nm and ~25 nm, respectively). The supercurrent oscillations in the resonator are excited by applying a microwave signal to the input, which is coupled to the resonator via the capacitive mirror. If desired, this signal can be a sum of two such waves (via a power combiner). The signal power that is transmitted through the resonator and escapes through the coupling capacitor at a given frequency is measured using an Agilent N5230A vector network analyzer (NA), on the output port of the resonator, after the signal has passed through a series of isolators, attenuators, and amplifiers. The transmission coefficient for this process, which we call the *S*-parameter, is defined via $S_{21} = 10\log_{10}\left(P_{NA}^{in}/P_{NA}^{out}\right)$ [or, equivalently, $S_{21} = P_{NA}^{in} - P_{NA}^{out}$, if $P_{NA}^{in}$ and $P_{NA}^{out}$ are expressed in dB], where $P_{NA}^{out}$ is the power of the signal sent *from* the NA *to* the resonator input, and $P_{NA}^{in}$ is the power measured *on* the NA input port which arrives *from* the resonator ouput port through a series of isolators, attenuators, and amplifiers. Isolators and attenuators adding up to ~30 dB on both the input line and output line from the resonator are inserted at low temperatures, in order to eliminate the thermal noise impact from the environment. A cryogenic low noise amplifier (from Low Noise Factory) and room-temperature amplifiers are also employed in order to increase the signal-to-noise ratio.

**3. One-wire case: Summary of experimental results**

For low values of the input signal power (e.g., around –60 dB), the transmitted power shows sharp, Lorentzian, resonance peaks centered at signal wavelengths, λ, obeying *L* = λ/2 (fundamental mode), λ (first harmonic), 3λ/2 (second harmonic),… , where *L* is the length of the resonator defined as the distance between the two mirrors (see Fig. 2a). As the power is increased towards –47 dBm, the frequency-dependent response near the λ/2 resonance first bends over towards lower frequencies, as shown in Fig. 2(b), doing so in a manner consistent with the behavior of the Duffing oscillator [16]. The nonlinear inductance of the nanowire, $L_{NW} = \frac{\hbar}{2e}\left(\frac{dI}{d\varphi}\right)^{-1}$, is the source of the Duffing nonlinearity, in which $e$ is the electronic charge, $\hbar$ is Planck's constant divided by $2\pi$, $I$ is the supercurrent, and $\varphi$ is the superconducting phase difference between the ends of the wire. As the power is increased further, the resonance then develops a marked *dip* in transmission near the center of the shifted peak (see Fig. 2b). In what follows, we refer to this dip as a *Lorentzian crater*. By contrast, we note that the resonance corresponding to the $\lambda$ mode remains Lorentzian up to much higher input powers (i.e., by a factor of ~3000). This difference in behavior is not accounted for by the difference in *Q*-factors (i.e., quality factors) between the resonances [32]. Indeed, the $3\lambda/2$ resonance, which is of still smaller *Q* than the $\lambda$ mode, develops a crater at a lower input power (see Fig 6a). The frequency width of the crater at this mode is larger than in the $\lambda/2$ mode because the crater width is inversely proportional to the quality factor. The crater width also grows with an increasing drive power. Thus, one is led to regard the crater as being related to the properties of the *nanowire itself*, and as being manifested at resonances that have an *antinode* at the location of the nanowire (the $\lambda/2$, $3\lambda/2$,…etc. modes) [16]. The properties of the film are expected to be observed at the $\lambda$, $2\lambda$,…etc. modes where there exists a node at the location of the nanowire. This view is corroborated by the relationship between the threshold power for crater formation, $P_c$, and the temperature (see Fig. 3), which has the form

$$P_c \propto \left[1-\left(T/T_c\right)^2\right]^3, \tag{1}$$

where $T_c$ is the critical temperature for the onset of superconductivity in the nanowire. Equation (1) matches the temperature dependence predicted by the Bardeen formula [33] for the *square* of a nanowire's critical current given by $I_c = I_{co}\left[1-\left(T/T_c\right)^2\right]^{3/2}$, where $I_{co}$ is the critical current at zero temperature and $T_c$ is the critical temperature of the nanowire. Thus, the crater is a result of dissipation triggered when the nanowire current exceeds it's critical current, resulting in Joule heating. The argument presented above is based on the assumption that, in general, the power carried through the resonator or a coplanar waveguide in general is proportional to the square of the current, i.e. $P \sim I^2$.

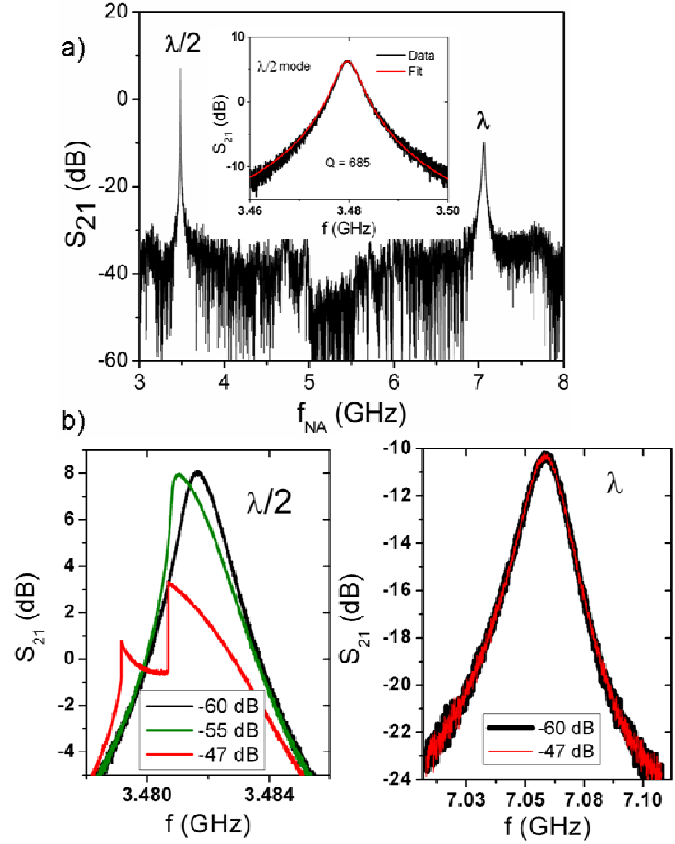

*Fig. 2.* (a) Transmission characteristic of sample S1, showing peaks at the fundamental frequency, corresponding to $L = \lambda/2$, and the first harmonic, at $L = \lambda$. Inset: the λ/2 transmission peak measured at low power with the corresponding Lorentzian lineshape fit using $Q = 685$. (b) (left) Shape of the λ/2 transmission peak for a relatively low power (black curve) and for a relatively high power (red curve). The peak becomes more asymmetrical and develops a "crater." (right) Shape of the λ peak ($Q = 335$) for the same powers as in the left panel; unlike the λ/2 peak, the λ peak shows no appreciable dependence on the input power in this regime. Note that the vertical axis is on a logarithmic scale, and that the quality factor is found by fitting the transmission curve to the Lorentzian lineshape: $S_{21} = \dfrac{A}{(f_\circ/Q)^2 + 4(f - f_\circ)^2}$, where $A$ is a scaling factor and $f_o$ is the resonance frequency.

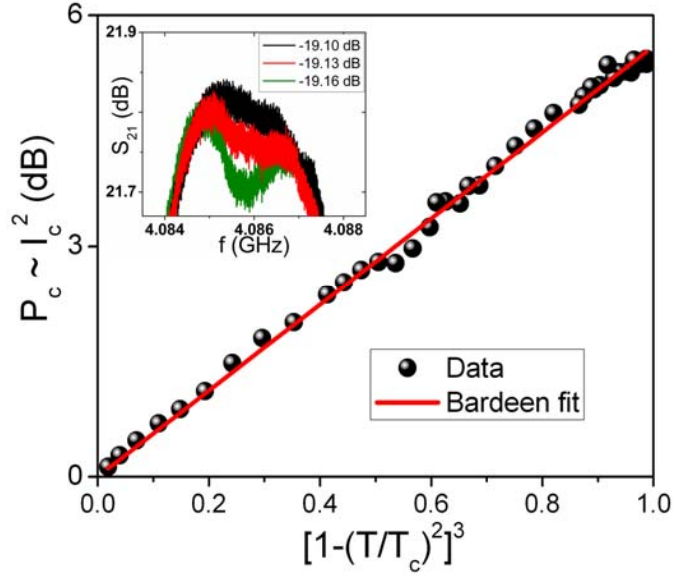

*Fig. 3*. Temperature dependence of the threshold input power, $P_c$, required for the onset of a crater for sample S2. That the onset power, $P_c$, is proportional to $\left[1-(T/T_c)^2\right]^3$ suggests that it is proportional to $I_c^2$, where $I_c$ is the critical current of the nanowires, and hence that the crater is a consequence of the nanowire current being driven past its critical value. Such a conclusion follows from the fact that according to Bardeen [33], the critical current of a nanowire depends on temperature as $\left[1-(T/T_c)^2\right]^3$, at all measured temperatures. In the fit, a value of 5.54 K was used for $T_c$, which is close to the $T_c$ values of other measured and similarly dimensioned nanowires. The quality factor of sample S2 at a temperature of 1.5 K was 725. The inset illustrates how the threshold power is determined: the red curve is taken to be the threshold, as it constitutes the power at which a crater just becomes observable. The uncertainty of the determined $P_c$ is about 0.05 dB.

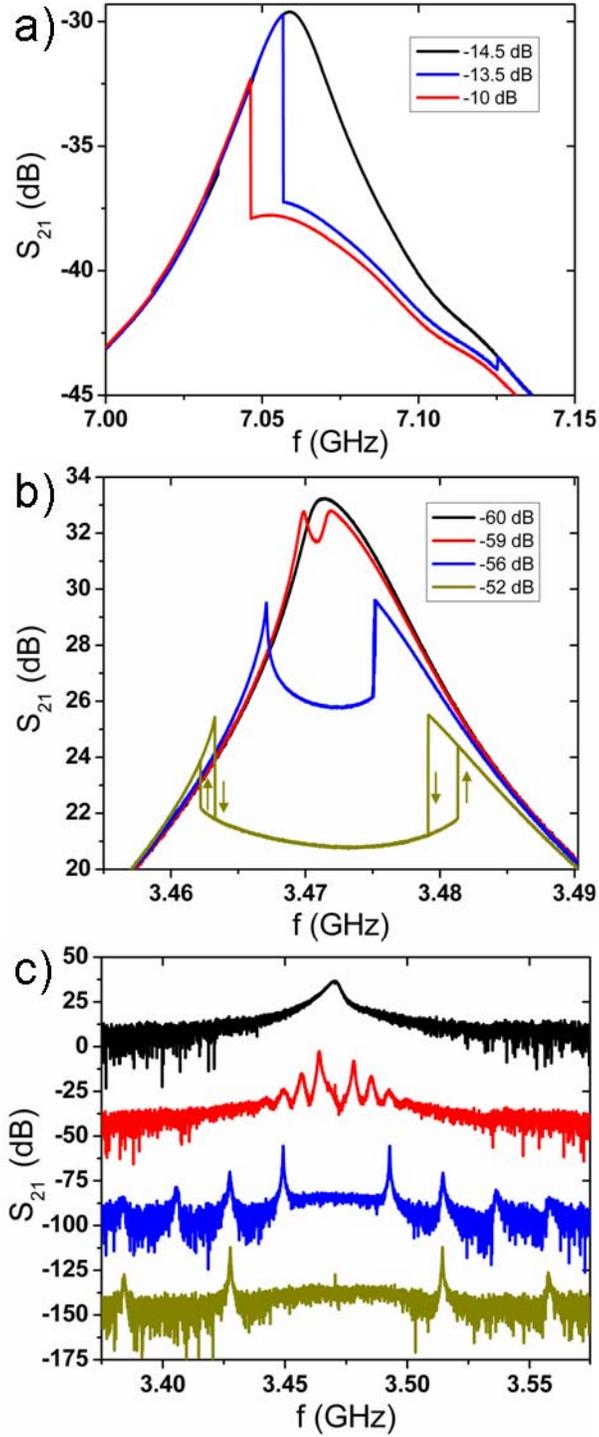

*Fig. 4.* (a) Transmission curve at the λ resonance as the input (i.e., drive) signal power is increased, showing the onset and expansion of the "crater" for sample S1. Note that there is no Duffing-type nonlinear behavior at this resonance, and that for powers above the threshold, the shape of the "crater" is concave-down.
(b) Evolution of the transmission curve at the λ/2 resonance as the input power is increased, showing the onset and expansion of the crater. Note that, for powers well above the threshold for crater formation, the transmission curve becomes hysteretic (i.e., irreversible with frequency sweeping). (c) The transmission spectrum [or, equivalently, in the time domain the supercurrent amplitude oscillations (see main text)] when the drive frequency is fixed at 3471 MHz; the four curves, from top to bottom, correspond to the input powers in panel (b). The curves in panel (c) have been translated vertically for ease of viewing.

Note that the positive curvature of the bottom of the crater at the $\lambda/2$ resonance (see Figs. 2b and 4b) is incompatible with a scenario (such as that presented for the simpler case of a resonator without any nanowires in Ref. [34]) in which the $Q$-factor of the resonance decreases abruptly at some critical input power, and the system thus enters a dissipative *stationary* state. The behavior at the $\lambda$ resonance is, however, consistent with this scenario (see Fig. 4a): this is to be expected because, for this resonance, the current amplitude is very small in the region where the nanowire is located, so that the nanowire dynamics do not participate strongly to this resonance. Under the simple $Q$-factor reduction scenario, the transmission coefficient would jump to a smaller value when the current exceeds the switching current (the value of the supercurrent at which the wire switches to the normal state) of the nanowire [35] and rise to its original level when the current falls below the retrapping current, but the lineshape would exhibit a negative curvature, as a perfect Lorentzian peak does. In other words, regardless of the $Q$-factor, there would be more transmission at the resonance than away from it.

A more striking inconsistency with the $Q$-factor reduction scenario is the occurrence of *current-amplitude oscillations* when the input power exceeds the threshold for crater formation. In this regime, the frequency spectrum of power transmitted by the resonator exhibits a periodic array of satellite peaks [see Fig. 4(c)], spaced at integer multiples of a certain frequency $\Delta f$ away from the drive frequency; $\Delta f$ increases with input power, and the height of these satellite peaks scales approximately as $1/n$, where $n$ is the $n^{th}$ peak, counting from the drive frequency. As the Fourier coefficients associated with a function that is periodic and has discontinuities decay as $1/n$ [36], the behavior of the satellite peak heights indicates that, in the *time* domain, the transmitted power exhibits periodic jumps.

For input powers near the threshold for crater formation, the crater is non-hysteretic; as the input power is increased further, however, hysteresis appears on the high-frequency side of the crater, and for still higher powers on the low-frequency side as well. Besides using a quasi-one-dimensional nanowire as opposed to a microbridge, the coexistence of hysteresis and amplitude oscillations is an important difference between our results and those presented in Ref. [29]. The hysteresis that we observe does not appear to be sensitive to the sweep rate (see Fig. 5).

Whereas nearly all of the data that we present is for a sample containing a MoGe nanowire of diameter 25 nm at the $\lambda/2$ resonance, we have observed essentially identical phenomena at the $3\lambda/2$ resonance (see Fig. 6), and also for even thinner nanowires. In addition, similar phenomena have been observed in resonators having much larger $Q$- factors, which incorporate much longer, thicker, and wider, Nb wires, where the width is much larger than the coherence length (Fig. 7).

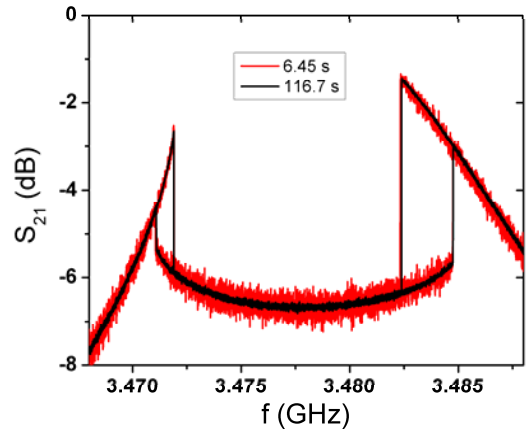

*Fig. 5.* Typical transmission characteristics, corresponding to two widely separated sweep speeds (red fast; black slow) for sample S1. The width of the hysteretic region (as well as the entire shape of the curve) does not depend noticeably on the sweep rate, at least for the sweep rates used in these measurements.

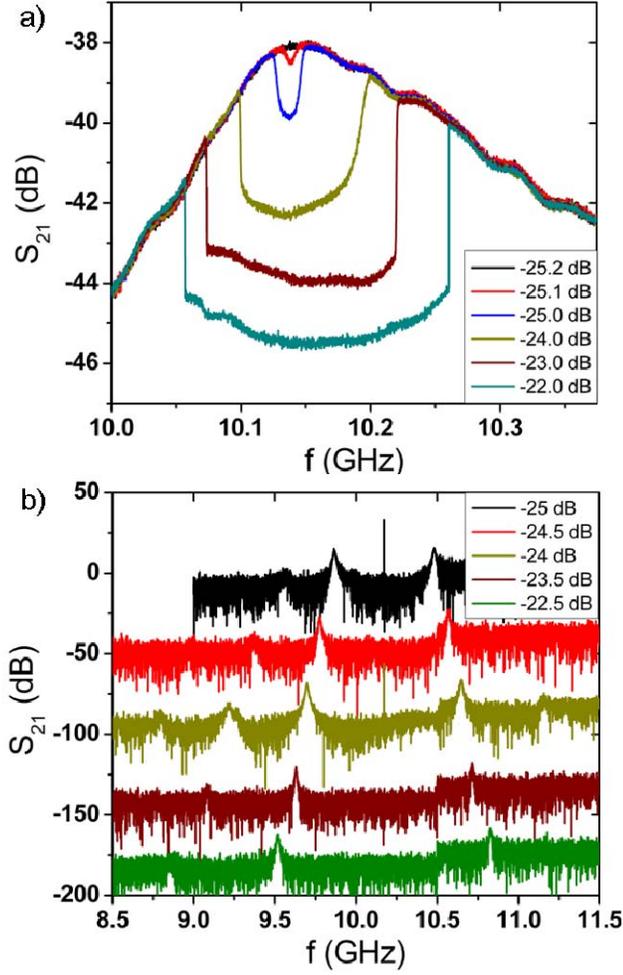

*Fig. 6.* (a) Transmission characteristics of sample S2 (25 nm thick MoGe) at 1.45 K at its $3\lambda/2$ resonance, as the input signal power is increased. In this case, the crater is qualitatively similar to that at the $\lambda/2$ resonance; however, it exhibits additional features, such as secondary dips near the edges of the crater. We attribute these new features to the fact that the $3\lambda/2$ resonance is relatively broad (*Q* ~60), and therefore requires higher input power for a crater to form. At these high input powers, the nonlinearities in the rest of the resonator (i.e., not the nanowire part), and the associated parasitic resonances, can no longer be neglected. (b) The satellite peaks when the driving frequency and power [which is in the same range as in panel (a)] resides in the crater (for the same sample S2). When the drive frequency is in the crater, the satellite peaks are qualitatively similar in shape to those at the $\lambda/2$ resonance. The curves in this panel have been vertically translated for ease of viewing.

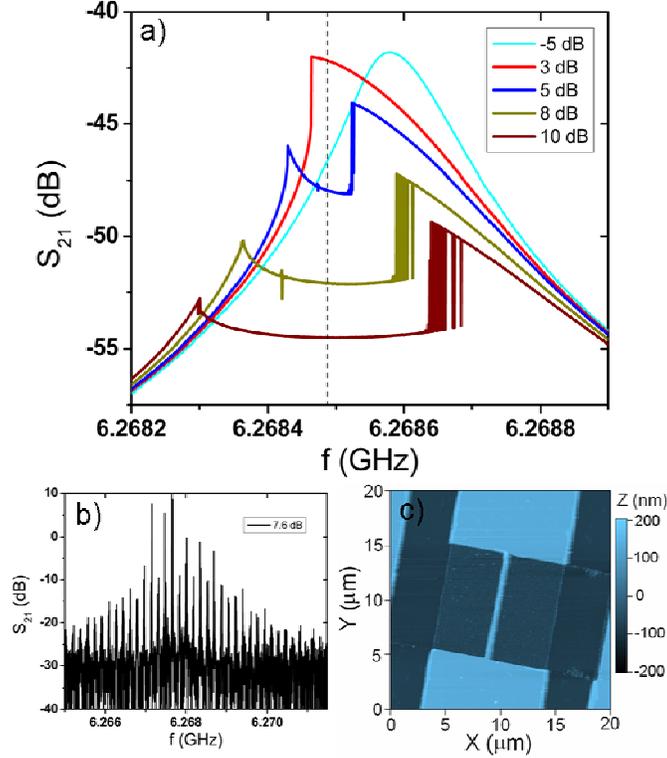

*Fig. 7.* Microwave measurements on sample S3, which is a Nb film resonator with an incorporated microbridge. The wire is 200 nm thick, 1 μm wide, and 10 μm in length and was fabricated using e-beam lithography. (a) Transmission characteristics of sample S3 at its λ/2 resonance, for a range of driving powers. At the measurement temperature of 1.5 K, the quality factor of the low-power curve is 43,500. The shape of the crater that forms is similar to that for the MoGe case. (b) Satellite peaks when the driving frequency and power lie inside the crater. The driving frequency is indicated by the vertical dashed line in panel (a), and has a power of 7.6 dB. (c) Superconducting Nb resonator, grown on a sapphire substrate and containing a microbridge.

## 4. Theoretical Model

*4.a Ingredients*

We have developed a phenomenological model for resonators with integrated nanowires, which may be summarized as follows. The resonator is modeled as a Duffing nonlinear oscillator [37] via the identification of the amplitude of oscillation with the supercurrent through the nanowire. We add to the model the following element, which we term a "switching rule": if the amplitude of oscillation exceeds a certain critical value (which corresponds to the switching current of the nanowire), the oscillation amplitude is instantaneously reset to zero. This switching rule is meant to capture the fact that when the current in the nanowire exceeds its switching current, the nanowire enters the normal state, in which it has a normal resistance of the order of 1 kΩ; while the nanowire is in its resistive state, the effective *Q*-factor of the resonator is very small; hence, the steady-state amplitude of the supercurrent is essentially zero. The switching rule represents the underlying assumption that as the wire switches to the normal state it will dissipate all the energy stored in the resonator (which is stored in two forms, namely the kinetic energy of the moving condensate and the potential energy of the electric field between the center conductor and the ground planes of the resonator). An additional element of the switching rule is that, after switching, the amplitude of the oscillator is held at zero for a fixed time $t_{hold}$, which corresponds to the time the nanowire takes to cool down and relax to its equilibrium superconducting state. (Alternatively, one can describe $t_{hold}$ as the duration for which the *Q*-factor of the resonator is taken to be zero.) At the end of $t_{hold}$, the *Q*-factor is returned to its original value, and the oscillation amplitude regrows according to the Duffing oscillator equation of motion.

An important aspect of the model, which is necessary for it to describe the observed hysteresis, is the continuous manner in which we take the drive frequency to be swept. This mimics the experimental situation, in which both the input current and the current in the resonator change *continuously*, as the network analyzer progresses from one value of the probe signal frequency to the next one. In particular, the initial conditions for the resonator at each drive frequency depend on the previous value of the drive frequency, and thus on the direction of the sweep. We must therefore ensure that the drive signal does not change discontinuously in our numerical simulation of the frequency sweep; as we shall see, this can be arranged by choosing the relative phase of the drive signal at frequencies $\omega$ and $\omega + \Delta\omega$ appropriately.

The algorithm outlined above is straightforward to implement numerically (for this we use the LabView environment), as we describe in Sec. 4c. As we shall see there, it yields results that fit our data very well, as shown in Figs. 10, 12, and 14. Before doing that, we explain in simple, physical terms why our model predicts the phenomenology that it does. For ease of presentation, we further simplify the model by neglecting the nonlinear character of the oscillator. The nonlinear element is responsible for the asymmetric nature of the crater, but is otherwise unrelated to the underlying physics. The nonlinear effects have been investigated elsewhere [16].

### *4.b Phenomenology*

*Basic picture of the oscillatory state*

The value $x(t)$ of the supercurrent in the resonator evolves in time *t* according to the oscillator equation:

$$\ddot{x}(t) + 2\kappa\dot{x}(t) + \omega_\circ^2 x(t) = Ve^{i\omega t}\Theta(t), \tag{2}$$

Where $\kappa$ is the damping coefficient, $\omega_\circ$ is the resonance frequency, and $Ve^{i\omega t}\Theta(t)$ is the driving signal, which as an amplitude $V$ and frequency $\omega$ and, as indicated by the $\Theta$ function, is switched on at time $t$. The physical current in the resonator is given by the real part of $x(t)$. When the instantaneous value of the supercurrent in the resonator exceeds the critical current of the nanowire, the following sequence of events occurs:

(1) The nanowire enters the normal state.
(2) The $Q$-factor, and, correspondingly the supercurrent in the resonator and voltage, both drop to zero, thus reducing the total stored energy to zero. All of these quantities remain zero for a time, which we denote $t_{\text{hold}}$.
(3) The $Q$-factor then returns to its equilibrium value (i.e., the value corresponding to small current-oscillation amplitudes), and the current begins to build up in the resonator, according to the oscillator equation.
(4) The current through the nanowire once again reaches its switching threshold, the nanowire switches, and the entire process repeats itself.

This cyclic process has a frequency $\Omega$, which is much lower than the resonance frequency $\omega$ of the oscillator. In other words, the frequency $\Omega$ is the frequency of the oscillations of the total amount of energy stored in the resonator and, at the same time, of the *amplitude* of the supercurrent oscillations. The time-dependence $x(t)$ of the displacement of the oscillator (i.e., the supercurrent in the resonator) is thus given by

$$x(t) = \cos(\omega t)\sum_{n=0}^{\infty} C_n \cos(n\Omega t) = \frac{1}{2}\sum_{n=0}^{\infty} C_n \left\{\cos\left[(\omega+n\Omega)t\right] + \cos\left[(\omega-n\Omega)t\right]\right\} \quad (3)$$

where $C_n$ is the $n^{\text{th}}$ Fourier component of the amplitude oscillations. Evidently, the Fourier transform of $x(t)$ consists of an array of regularly spaced spikes, offset from the central frequency $\omega$ by integer multiples of the amplitude oscillation frequency $\Omega$. The model thus readily explains the experimental fact that the $n^{\text{th}}$ spike is $\sim 1/n$ times the height of the first spike (for $n \geq 1$)– this corresponds to the asymptotic power-law decay behavior of the set of Fourier coefficients, discussed above.

*Form of the crater*

We now turn to step (3) of the amplitude oscillation cycle, viz., the growth of current in the resonator. Once again, we treat the instructive case of a damped, driven *linear harmonic* oscillator, governed by the oscillator equation

$$\ddot{x}(t) + 2\kappa \dot{x}(t) + \omega_\circ^2 x(t) = Ve^{i\omega t}\Theta(t). \quad (4)$$

Here, the step function $\Theta(t)$ reflects the fact that the drive begins abruptly at the end of $t_{\text{hold}}$. (Equivalently, and more physically, we could have attached the step function to the inverse of the dissipative term; however, the present form is simpler to analyze.) The general solution for $x(t)$ may be written in terms of the corresponding Green function $G(t,t')$ as follows:

$$x(t) = \int_{-\infty}^{t} dt' G(t,t') V e^{i\omega t'} \Theta(t') = \int_{0}^{t} dt' G(t,t') V e^{i\omega t'}, \tag{5}$$

where $G$ is given (for $t > t'$) by

$$G(t,t') = \frac{1}{2\sqrt{\omega_\circ^2 - \kappa^2}} \left[ e^{-\kappa(t-t')} \sin(\omega_1(t-t')) \right], \tag{6}$$

in which

$$\omega_1 \equiv \sqrt{\omega_\circ^2 - \kappa^2} \tag{7}$$

is the shifted resonance frequency. Upon performing the relevant integration, one finds that the current is given by

$$x(t) = \frac{1}{2i\omega_1} \left( \frac{e^{i\omega t} - e^{-\kappa t} e^{i\omega_1 t}}{\omega - \omega_1 + i\kappa} \right) - \text{r.r.} \tag{8}$$

Note that we have omitted the much smaller, rapidly rotating (denoted r.r. and on the order of GHz), terms, as our primary focus is on the shape of the *envelope* of the current, i.e., the manner in which the amplitude grows. Note that $x(t)$ vanishes at $t = 0$, as desired, and that eventually the amplitude saturates to its steady-state sinusoidal form, doing so on a timescale given by $1/\kappa$. In Fig. 8 we show the behavior of $x(t)$ at fixed $\kappa$ for various values of the detuning (with the aforementioned rapidly oscillating terms omitted), which is given by: $|\omega - \omega_\circ|$.

We now proceed to explain how the nature of the process by which the current grows explains both the concave-up form of the crater and the existence of hysteresis.

*Shape of the crater*

The scenario outlined in the previous subsection offers a natural explanation for why the craters predicted by the model are concave *up* rather than concave *down*. From the model, and specifically Eqn. (8), we know that (a) the r.m.s. value of $x(t)$ grows linearly, for $t \ll 1/\kappa$, with a frequency-independent slope, and (b) the second derivative of the r.m.s. value of $x(t)$ is *negative* for times until the point at which $x$ saturates at its maximum value $x_{max}$, which must be greater than some critical value $x_c$ if a crater is to form. [The two typical forms of $x$ are shown in the insets of Fig. 8.]

It follows from point (a) that, if $x$ reaches $x_c$ at a very early time (i.e., if the drive power is very large) then the r.m.s. value of $x(t)$ (which we term $\bar{x}$ and define to be the mean taken over the interval $2\pi/\Omega$, i.e., over one cycle of the amplitude oscillation) can be arrived at by expanding the exponential to first order in $t$:

$$\bar{x} = \frac{x_c}{2}\left(\frac{2\omega_1 x_c/V}{t_{hold}+(2\omega_1 x_c/V)}\right). \tag{9}$$

On the other hand, it follows from point (b) that, as the drive frequency is swept away from resonance, $x(t)$ takes a relatively long time to reach $x_c$ (i.e., when $x_{max}$ is relatively close to $x_c$, as it would be at driving frequencies relatively far from the resonance) and the average transmission *increases*. This occurs for two reasons: (i) as the amplitude of $x(t)$ saturates toward $x_{max}$, a larger part of each cycle is spent at higher supercurrent amplitudes [by observation (b); see also Fig. 11(c) and (d)]; and (ii) the amplitude oscillation period $2\pi/\Omega$ increases, and therefore $t_{hold}$ occupies a smaller part of the cycle resulting in an increase in transmission. Thus, as one moves away from the resonance, $x_{max}$ decreases, the transmitted power increases, and consequently the crater is concave up.

*Hysteresis*

Depending on whether or not the detuning exceeds $\kappa$, the oscillation amplitude either (i) overshoots its steady-state value or (ii) does not. For instance, if the drive is exactly on resonance, we know from Eqn. (8) that the envelope behavior of the current amplitude $A(t)$ of takes the form

$$A(t) \sim 1 - e^{-\kappa t}, \tag{10}$$

and thus approaches its steady-state value monotonically. By contrast, if the detuning is large compared with $\kappa$, the envelope behavior of the supercurrent amplitude exhibits beats, i.e.,

$$A(t) \sim \sin\left[(\omega-\omega_1)t\right], \tag{11}$$

and can overshoot its steady-state value by up to a factor of two. [Note that if the drive were turned on adiabatically (or, equivalently, the $Q$-factor were decreased adiabatically), the amplitude would achieve its steady-state value without overshooting. In this case there would be no hysteresis. Therefore, the point at which switching occurs depends on whether the current is increasing rapidly or adiabatically; as we shall see, this leads to hysteresis.]

For large drive strengths as the frequency is swept, $x$ first reaches its critical value relatively *far* from resonance, and case (i) applies. In this case, $x_{max}$ exceeds the *steady-state* amplitude $x_{ss}$. If $x_c$ lies between $x_{ss}$ and $x_{max}$ [see Fig. 8(b)], the model predicts that bistability occurs—if the system is initialized in its steady state, it can stably continue in the steady-state; however, once the system switches, it cannot re-enter the steady state because to do so it would have to go through the entire transient, which overshoots $x_c$. Therefore, the sweep *toward* resonance (during which the amplitude adiabatically increases and the resonator enters the bistable region initialized in the steady state) differs from the sweep *away* from resonance (during which the system is initialized in the oscillatory or pulsing regime, and cannot reach the steady state), and the transmission curve thus exhibits hysteresis.

By contrast, for relatively small drive powers, such as those that are barely sufficient to generate a crater, $x$ reaches $x_c$ at frequencies less than $\kappa$ from the resonance. Now it is case (ii) that is realized, and therefore $x$ is in the monotonic-growth regime, in which $x_{max} = x_{ss}$, and the model therefore predicts that no hysteresis should occur—as found in the data.

We conclude this heuristic discussion with some brief remarks on how the inclusion of thermal or quantum fluctuations, or the Duffing nonlinearity, would affect the foregoing arguments. One would expect fluctuations of the photon number in the resonator to cause fluctuations in the current through the nanowire; and these would sporadically drive the current across $x_c$, and should therefore cause switching between the two stable states in the hysteretic region [see Fig. 7(a) and Fig. 12(a)]; at sufficiently high temperatures this would lead to the disappearance of hysteresis. As for the Duffing nonlinearity, which must be incorporated to achieve *quantitative* agreement with the data, it does not *qualitatively* affect the above considerations: it causes the two sides of the resonance curves to have distinct hysteretic behavior and forms of crater, but each side would still individually behave essentially as one would predict using the linear-oscillator model.

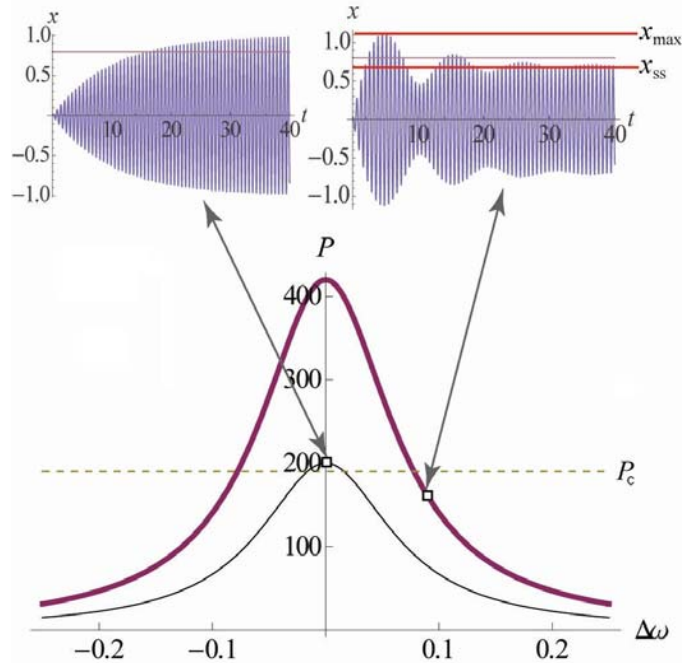

*Fig. 8.* Time averaged transmission characteristic curves (main panel) and the growth of transients (insets) computed for a linear oscillator. The lower panel shows the transmitted power as a function of input frequency for a linear

oscillator, for two input-powers; the horizontal line indicates the power required to form a crater. For low input power, the onset of the crater lies near resonance; in this regime, the growth of transients is monotonic, as shown in the upper left panel, and (as explained in the text) no hysteresis occurs. For high input power, the onset of the crater is far from resonance; in this regime, the growth of transients is *non-monotonic*, as shown in the upper right panel. Hence, the maximal transient amplitude $x_{max}$ exceeds the steady-state amplitude $x_{ss}$, and thus bistability arises, as explained in the main text. The thin, horizontal lines in the two upper panels correspond to the critical current of the nanowire; the thick lines correspond to $x_{max}$ and $x_{ss}$ as defined in the text.

### *4.c Fits to data*

We implement the model using a LabVIEW program that solves the circuit sketched in Fig. 9, which uses the Josephson junction inductance as the nonlinear Duffing element. The three circuit parameters used to fit the data—viz., the capacitance, inductance, and resistance—are obtained by fitting the low-power, near-Lorentzian resonance. We model the nanowire as an *effective* Josephson junction having a switching current $I_{sw}$ and kinetic inductance $L_k$ that is determined, as discussed below, by fitting the extent to which the resonance is non-Lorentzian at the onset of the crater. A timestep of 1 ps is used to advance the computation. Using much smaller timesteps results in inaccurate low-power simulations, due to the large periods of the supercurrent oscillations and a loss in accuracy of the order parameter phase due to precision limits within the program. Also, using much larger timesteps results in the breakdown of the approximation used to advance the iterative method in solving the differential circuit equation. The value for the $S_{21}$ parameter is calculated at each particular frequency as
$S_{21} = 10 \log[\langle I_s \rangle^2 / I_b^2] + \text{offset}$, where $I_s$ is the supercurrent in the resonator, $I_b$ is the bias current applied to the circuit in Fig. 9, and the offset is used to account for the reference value of $S_{21}$ effected by the combination of attenuators, isolators, and amplifiers.

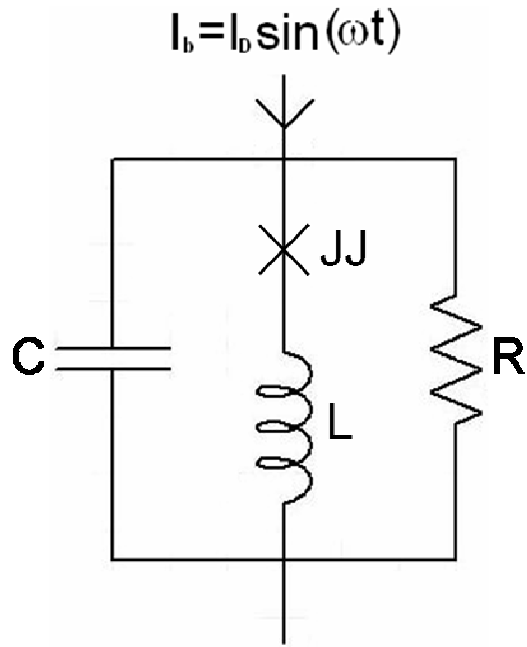

*Fig. 9.* Schematic depiction of the circuit diagram used to model the spectrum of transmission curves. The diagram consists of a resistively and capactively shunted junction (RCSJ) circuit having an inductor inserted in series with the nanowire. The circuit is driven by a sinusoidal bias-current of amplitude $I_D$ and frequency $\omega$.

The model described in Secs. 4(a) and (b), if we augment the left-hand side of the oscillator equation with the nonlinearity arising from the wire (as explained below), is able to *quantitatively* reproduce the data taken at various temperatures. In particular at a temperature of 1.5 K, as shown in Figs. 10, 11, and 12, the model quantitatively reproduces the following features: (1) the evolution of the crater shape, as the input power is increased; (2) the dependence of the satellite-peak spacing $\Delta f$ on the input power at a fixed frequency; and (3) the dependence of the satellite-peak spacing on input frequency at a fixed power.

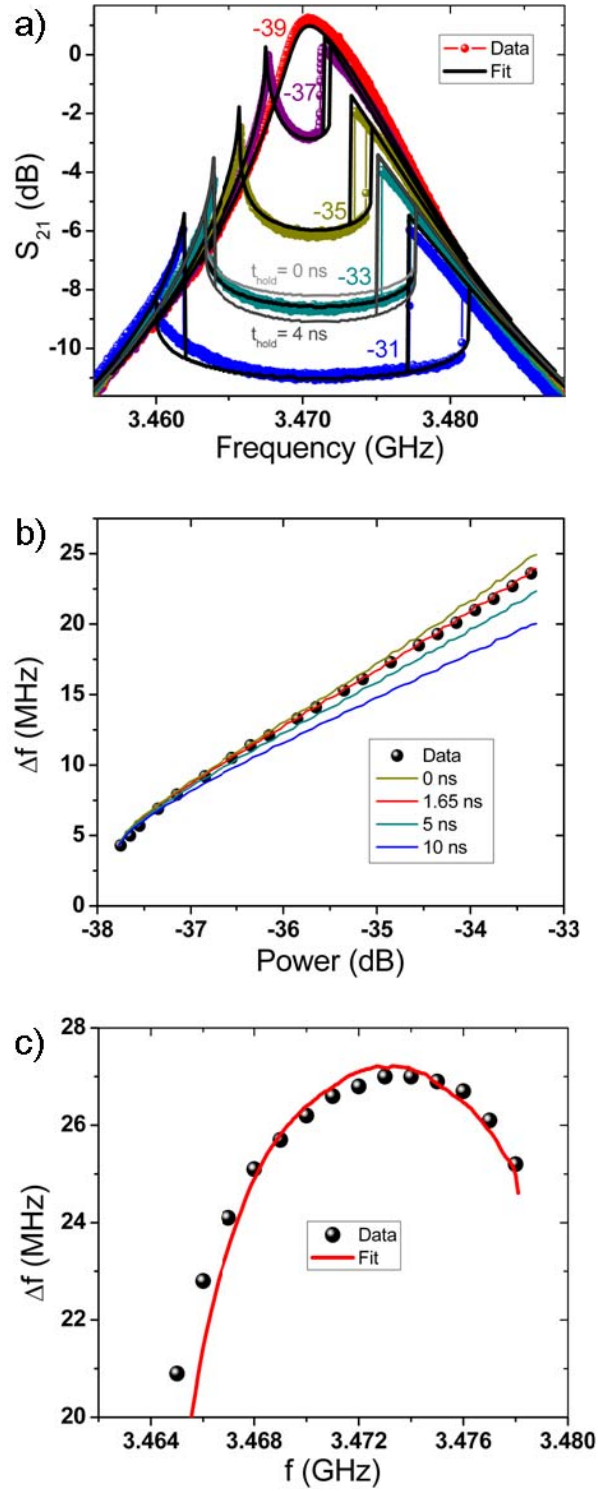

*Fig. 10.* Comparison of experimental data for sample S1 and predictions of the model at $T = 1.5$ K. (a) Transmission characteristics for input powers in 2 dB increments (thick colored lines), and fits to the model (thin black lines). At an input power of -33 dB, two other simulations are shown in grey and dark grey corresponding to $t_{hold} = 0$ and $t_{hold} = 4$ ns, respectively. The crater grows deeper

(shallower) as the $t_{hold}$ parameter is increased (decreased).  Thus, $t_{hold}$ is a sensitive and necessary fitting parameter that only affects the depth of the crater in this graph.  Here $t_{hold}$ = 1.65 ns results in the best fit. (b) Satellite peak spacing ($\Delta f$) vs. input-power at a fixed input-frequency of 3471 MHz. (c) Satellite peak spacing ($\Delta f$) vs. input frequency at a fixed input power of -33.6 dB.  All fits were calculated using the following fitting parameters:  $C$ = 16.7 pF, $L$ = 0.113 nH, $R$ = 995 Ω, $L_k$ = 13.1 pH, $I_{sw}$ = 8.98 μA, and $t_{hold}$ = 1.65 ns.

The supercurrent, which is given by $I_s = I_c \sin(\phi)$, was calculated as a function of time by numerically integrating the circuit equation for Fig. 9 to evolve the phase across the junction at each time step according to

$$\ddot{\phi} = \left[ \frac{I_{cap}/C + I_c L \dot{\phi}^2 \sin(\phi)}{\hbar/2e + I_c L \cos(\phi)} \right]. \tag{12}$$

This equation was derived by equating the voltages on the capacitive and junction branches of the circuit given in Fig. 9 and solving for $\ddot{\phi}$.  Once $\ddot{\phi}$ is known, Kirchhoff's current law can be used to solve for $I_{cap}$ to advance the computation as follows $I_{cap} = I_b - I_s - I_R$, where each term on the right hand side is known given the value of $\phi$ or one of its derivatives from the previous timestep.  The superconducting phase $\phi$ and all its derivatives are initialized to zero.

In a frequency and power regime outside the crater and near to the resonance, the model shows the supercurrent monotonically growing [Fig. 11(a)] as expected from Eqn. (10); whereas far from the crater, the model reveals the nonmonotonic growth analogous to that in Fig. 8 as expected from Eqn. (11) [Fig. 11(b)].  Inside the crater regime, the supercurrent grows monotonically; even far from the resonance there is monotonic growth, due to the fact that the supercurrent reaches its maximum amplitude before another period in its oscillatory behavior is reached when the transient would overshoot and cause nonmonotonic growth.  Thus, monotonicity is enforced inside the crater.  The role of $t_{hold}$ can also be clearly be visualized in Fig. 11(c) and (d): it corresponds to the interval in which the supercurrent is held at zero after the nanowire switches to the normal state.  Once this time is over, the supercurrent begins to grow.  Additionally, the satellite peak frequency spacing $\Delta f$ can be obtained from the model, which are calculated by taking the Fourier transform of the supercurrent versus time profile.  These fit the data well, as can be observed in Fig. 11(c) and (d).

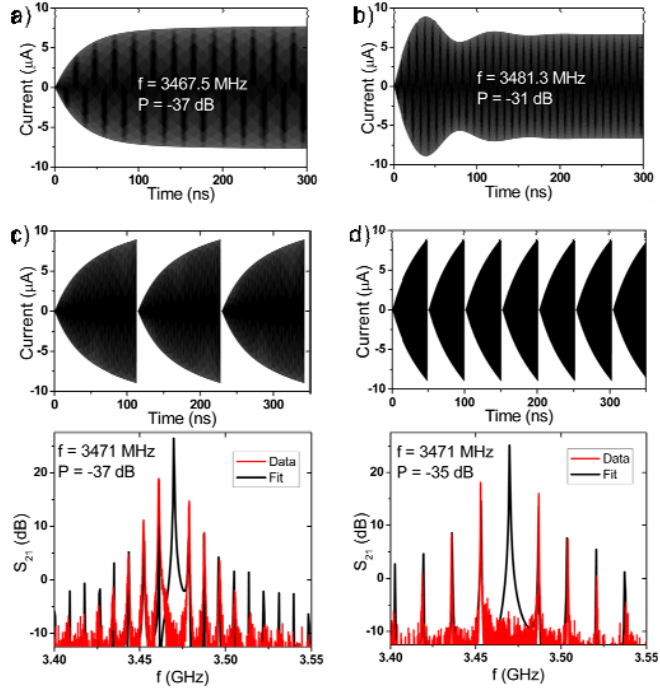

*Fig. 11.* Model prediction of supercurrent growth in the resonator, and comparison between satellite peak data and model for two powers for sample S1. (a) Supercurrent growth in the resonator at a power of -37 dB and frequency of 3467.5 MHz, which is near the resonance and outside, but near the crater shown in Fig. 10(a). The supercurrent growth is monotonic in this regime. (b) Supercurrent growth in the resonator at a power of -31 dB and frequency of 3481.3 MHz, which is far from resonance and also outside, but near the crater shown in Fig. 10(a). (c) The upper panel shows the supercurrent growth in the resonator inside the crater at a drive power of -37 dB and a frequency of 3471 MHz. The lower panel shows the satellite peaks (i.e., the Fourier transform of the upper panel) at this drive power and frequency and the corresponding fits to the model. (d) Same as (c), except for a drive power of -35 dB. As the drive power is increased, the satellite peak spacing increases, as the model predicts. The central peak in (c) and (d) is too narrow to be observed in the data.

As discussed in section 4(b), at finite temperatures, thermal fluctuations can reduce the hysteresis by causing switching between the two stable states. The difference in thresholds the system would exhibit with and without the presence of significant thermal (or quantum) fluctuations on the sweep towards resonance would be greater than on the sweep away from resonance. This can be seen as follows: on the sweep toward resonance, the current grows adiabatically to its critical value, as described in Sec. 4b, whereas on the sweep away from resonance, the current grows rapidly to its critical value. In the former case, the current amplitude is constantly near its critical value; therefore, any fluctuation will carry it past this value and cause the system to enter the crater. In the latter case, only fluctuations that occur during the brief interval that the current is near-critical will have an effect. This effect can be seen in Fig. 12(a), where for a power of -41 dB, the simulation is shown for both the case of including and not including thermal fluctuations. In these fits, thermal noise can be included by adding a random number with a given amplitude (here, $1.91\times10^{-4}$) to the phase at each time step. The choice of the phase fluctuation amplitude can then be checked by calculating the resulting supercurrent fluctuations $I_{fluct}$ predicted by the model and comparing it to the estimate from the equipartition theorem: $\frac{1}{2}LI_{fluct}^2 = \frac{1}{2}k_BT$, where $L$ is the total inductance of the resonator-nanowire system. At a temperature of 1.5 K and with an inductance of ~ 0.2 nH, the supercurrent fluctuation in the resonator can be estimated from the equipartition theorem to be $I_{fluct} \approx 300$ nA, which matches the modeled supercurrent fluctuations. This agreement of the predicted fluctuations and the fluctuations needed to produce the best fits confirms that thermal fluctuations are responsible for the observed small value of the hysteresis in this sample with a relatively low critical current.

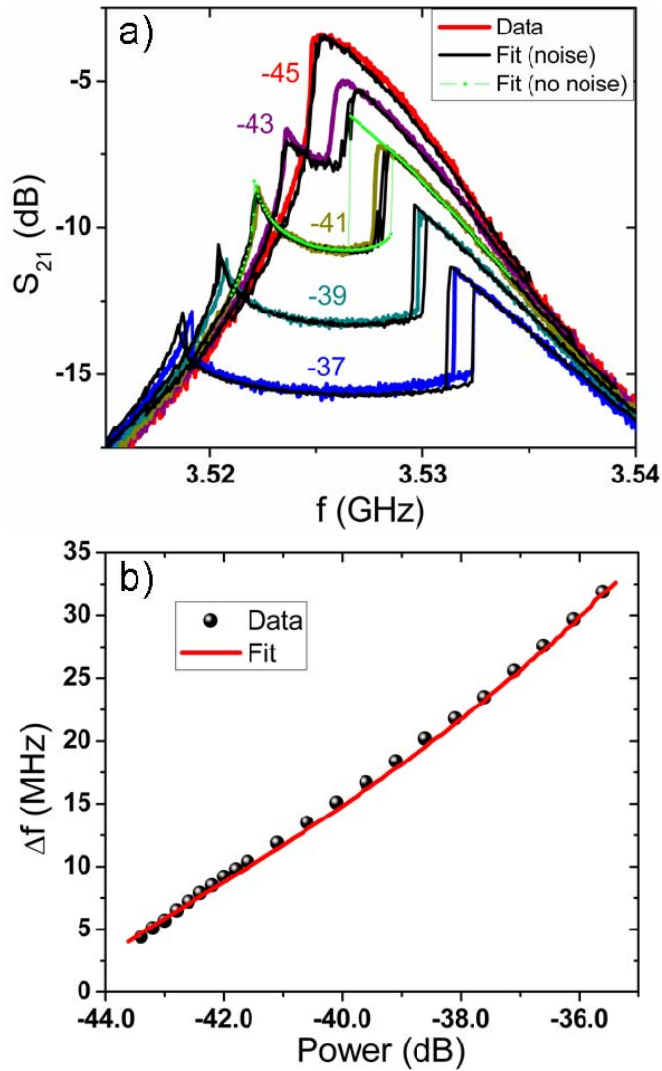

*Fig. 12.* Comparison of experimental data for sample S4, which is a 10 nm thick MoGe resonator/nanowire, and predictions of the model at 1.5 K.
(a) Transmission characteristics for input powers in 2 dB increments (colored lines), and fits to the model (black lines). The bright green curve at −41 dB shows the model simulation upon excluding phase noise. The quality factor is 665. (b) Satellite peak spacing vs. input power at a fixed frequency of 3525 MHz. The data and model show good agreement. All fits were calculated with the following fitting parameters: $C = 7.55$ pF, $L = 0.204$ nH, $R = 3.2$ k$\Omega$, $L_k = 65.4$ pH, $I_{sw} = 1.584$ μA, $t_{hold} = 1.8$ ns, and $R_{th} = 0.38$ m$\Omega$.

The data at much lower temperatures, such as 300 mK, deviate slightly from the model's predictions, in that the crater has a pronounced left-to-right gradient, as shown in Fig. 13(b). The origin of this effect is not clear; however, subtracting off a linear term (having a coefficient of 87 ndB/Hz) from *all* the resonance data is sufficient to bring the data into good agreement with our model, as shown in Fig. 13(a). We therefore believe that this slope is extrinsic to the properties of the nanowire, and is due, instead, to the low-temperature behavior of the two-dimensional parts of the resonator or to the other circuit elements, or due to parasitic coupling through the vacuum.

*Interpretation of fit parameters*

In the fits to the data shown in Figs. 10(a), 12(a), and 13(a), the Duffing oscillator parameters are determined by fitting the subcritical (i.e., craterless) resonance data. There are two further fitting parameters: (1) the drive power $P_c$ required for the onset of the crater; and (2) the interval $t_{hold}$, for which the resonator is taken to be quiescent, once the nanowire enters its normal state. As discussed in Sec. 3, our identification of $P_c$ with the power at which the current through the nanowire reaches $I_c$ is supported by the temperature dependence of $P_c$. In principle, $I_c$ is deducible from the coefficient of the Duffing term, using the current-phase relation (CPR) of the nanowire: however, the CPR appropriate for MoGe nanowires has not been well characterized, to date; we have therefore found it more reliable to determine the coefficient of the nonlinearity experimentally.

*Implications for relaxation phenomena in the nanowire*

Our other fit parameter, $t_{hold}$, is sensitive to relaxation processes in the nanowire. *Prima facie*, it might seem that $t_{hold}$ should depend on the *longer* of the following intervals: the timescale on which the current in the resonator relaxes to its dissipative steady-state value, and the timescale on which enough heat flows out of the nanowire that it can re-enter the superconducting state after it enters the normal state due to heating. As we can infer from the inductance and normal-state resistance of the nanowire (which is on the order of ~ 100 pH and 1 k$\Omega$, respectively), the former interval is too short to explain the measured value, which can be estimated by $L/R = 0.1$ ps; besides, in order to fit the data within our model it is natural to assume that the current in the circuit *goes to zero*, which would not be the case if the nanowire were to re-enter its superconducting state before the resonator had relaxed. Thus, we can assume that $t_{hold}$ depends on the relaxation rate of the nanowire back into the superconducting state. This being so, one might expect $t_{hold}$ to depend strongly on the temperature of the leads (i.e., the bath temperature), as the thermal conductivity of a gapped BCS superconductor decreases exponentially at low temperatures. In fact, however, $t_{hold}$ *does not* seem to depend appreciably on the bath temperature at low bath temperatures (i.e. in the range 0.3-1 K), as can be seen from Fig. 13(b). (At bath temperatures higher than 2 K, however, our fits find $t_{hold}$ to be zero, to within our uncertainty.)

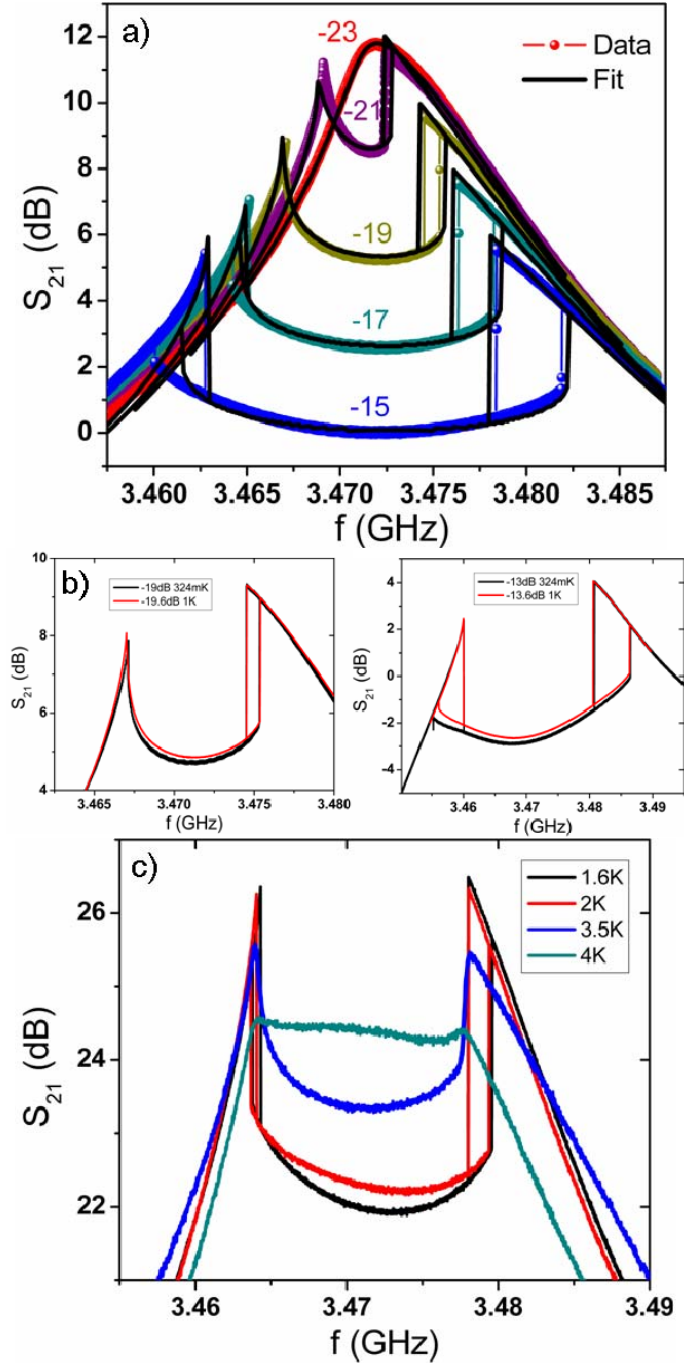

*Fig. 13.* (a) Transmission characteristics of sample S1 at 324 mK for input powers in 2 dB increments (thick colored lines), and fits to the model (thin black lines). At this temperature the thermal fluctuations are negligible and were not included in the modeling. At low power, $Q = 425$. The fits were calculated using the following fitting parameters: $C = 15.2$ pF, $L = 0.125$ nH, $R = 1005$ Ω, $I_{cL} = 25.14$ µA, $I_c = 10.39$ µA, and $t_{hold} = 2.9$ ns. (b) Dependence of crater depth on temperature for craters of fixed width (for two different widths). Within our model, this depth should be sensitive only to the fit parameter $t_{hold}$; the very weak

temperature dependence of the crater depth indicates that $t_{hold}$ does not depend strongly on temperature, at least over the range 300 mK to 1 K. Two features are of note here: (i) $t_{hold}$ is slightly longer for lower temperatures; and (ii) as the model predicts, the crater depths at higher input power are more sensitive to the value of $t_{hold}$. (c) Craters at various temperatures that exhibit similar crater widths. At lower temperature the crater can be fit with the model including a nonzero $t_{hold}$. At slightly higher temperature, $t_{hold}$ begins to decrease towards zero. At still higher temperatures, the current noise in the system becomes comparable to the signal current and the crater becomes flat. Each graph was horizontally (by ~ 100 KHz) and vertically (by ~ 0.5 dB) translated (to compensate for the temperature-dependence of the resonance frequency and other parasitic effects) for easier crater-width comparison.

Thus, taken together, our measurements and modeling lead us to the perhaps surprising conclusion that at low enough temperatures the time it takes the nanowire to relax back into the superconducting state, in the absence of a current, does not depend strongly on the bath temperature. A possible scenario that is consistent with this observation goes as follows: in the middle of the nanowire the superconducting gap collapses and reforms essentially immediately (i.e., on the Ginzburg-Landau timescale). This does not give the normal electrons and holes created during the collapse of the gap sufficient time to equilibrate; thus, located near the center of the wire are a substantial number of Bogoliubov quasiparticles, having energies comparable to the gap energy. Within this scenario, the number of quasiparticles created depends only on the highest temperature achieved by the nanowire during the collapse process, and is therefore essentially independent of the bath temperature. Relaxation of the nanowire occurs via the diffusion of these quasiparticles into the leads; this process occurs at a rate that depends on the *effective mass* of the Bogoliubov quasiparticles, which is proportional to the gap, as illustrated in Fig. 14. However, the magnitude of the gap saturates at low temperature, and therefore so does the rate of diffusion of quasiparticles. Thus, the scenario outlined above would suggest that $t_{hold}$ should saturate at low temperatures, as we observe experimentally.

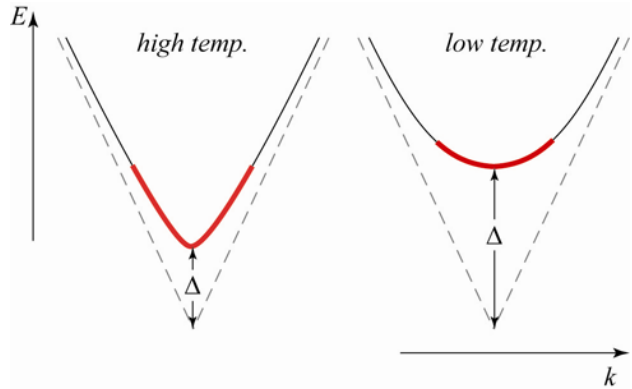

*Fig. 14.* Dispersion relations of trapped Bogoliubov quasiparticles expected under the scenario outlined in the present section, at high and low bath-temperatures. The chief difference between the two cases is the value of the superconducting gap $\Delta$, which is larger at low temperatures. The average quasiparticle velocity (given by $\partial E / \partial(\hbar k)$) decreases as the gap increases, which can be observed in the *k*-space denoted by the thick red line; hence, the rate at which quasiparticles diffuse into the leads also decreases. However, this diffusion rate, along with the gap, is essentially constant at low temperatures.

**5. Resonators bridged by two nanowires**

*5.a. Motivation and background*

We now turn to the case of resonators that incorporate two wires, as shown in Figs. 1(c) and (d). These devices are similar in some respects to the resonators interrupted by SQUIDs that were studied in Ref. [38, 39]; they differ, however, in two crucial respects. (i) Nanowires, unlike Josephson junctions, can support metastable states having one, or many, virtual vortices trapped in the region between the wires; these devices may therefore enable the study of, e.g., the quantum tunneling of vortices across the nanowires. (ii) If the entire device—resonator, including nanowires—is placed in a perpendicular magnetic field, various properties of the system, such as the intrinsic resistance and the switching current, are periodic in the magnetic field; however, the period is much shorter than what can be estimated by dividing the flux quantum by the area of the loop formed by the nanowires (or, equivalently, the conventional Little-Parks geometrical area dictated value). Suppose that the wires are separated by a distance *a*, and is each of length *b*, so that the area between the wires is given by *ab*: for the *conventional* Little-Parks effect, the properties of the wires should oscillate with magnetic field with a period given by $\Delta B = \Phi_0 / ab$, where $\Phi_0$ is the superconducting flux quantum. This effect is, however, greatly modified in situations such as the present one, in which the leads are themselves mesoscopic—i.e., have widths smaller than the London penetration depth $\lambda$. In such cases, it can be shown [30, 40] that the magnetic-field periodicity of the properties of the system as a whole is largely set by screening currents *in the leads*, which do not depend on the length of the wires. Thus, e.g., if the width *l* of the leads is much greater than *a*, the effective periodicity of the physical properties of the wires for small fields is given not by $\Delta B = \Phi_0 / ab$, but by $\Delta B = \Phi_0 / c_1 al$, where $c_1$ is a geometry-dependent number of order unity. At stronger magnetic fields, vortices enter the leads, and the periodicity changes.

The effects discussed in the preceding paragraph have been explored both experimentally and theoretically for the case of d.c. currents [30, 31, 40]. The experiments reported here confirm that the Lorentzian crater is also periodic in the magnetic field, with the same periodicity.

*Magnetic-field dependence of resonance and Lorentzian crater*

Given our previous association—discussed in Sec. 3—of the power required for the onset of the crater with the switching current of the nanowire, one would expect that the power required for the onset of the crater in the two-nanowire situation would depend periodically on the applied magnetic field. We do indeed find such a dependence (Fig.

15), with a periodicity consistent with that predicted in [30, 31, 40]. The critical power $P_c$ at the onset of the crater is obtained using the $S_{21}$-parameter in the dB scale: $P_c = S_{21} + P_{NA}^{out}$. The theoretical period of the magnetic field, for which the sample is tilted at an angle, $\theta$, with respect to the perpendicular magnetic field in order to fit in our measurement system, is calculated from: $\Delta B = \Phi_0 / c_1 al \sin|\theta|$. A wire separation of 6.63 µm, a lead width of 10 µm, and an approximate tilt angle of 35-40° results in a theoretical prediction of $\Delta B$ = 48.5 to 54.4 µT, assuming $c_1 \approx 1$. This is close to the experimentally measured vale of $\Delta B$ = 41.4 µT. The small difference in the predicted magnetic field period can be accounted for through the geometic parameter $c_1$, and through the uncertainty of our knowledge of the exact angle between the resonator surface and the applied magnetic filed (the sample was not horizontal due to practical limitations related to the dimensions of the sample holder and the cryostat).

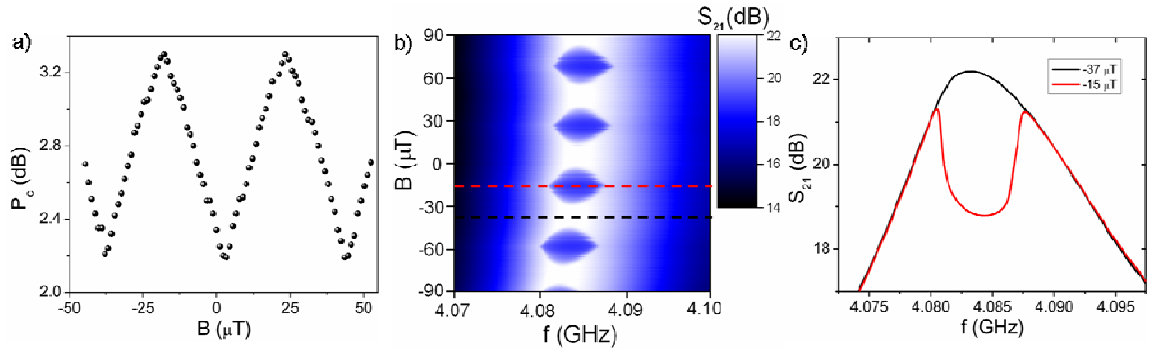

*Fig. 15.* (a) Oscillations of crater onset power $P_c$ as a function of magnetic field for sample S5, which is a 25 nm thick MoGe resonator containing two relatively symmetrical nanowires (having similar critical currents) separated by 6.63 µm and connecting a center conductor that is 10 µm in width. (b) Map of the transmission coefficient as a function of magnetic field and drive frequency, at fixed drive power, showing the periodic onset and disappearance of the Lorentzian crater (i.e., the dark islands in the middle). Regions of higher transmission are shaded more lightly. (c) Two transmission curves for the same input power but at differing magnetic fields. The drive power in this case lies between the minimum and maximum drive powers for magnetic field dependent crater onset. The black (red) curve in panel (c) is shown in panel (b) as a dashed black (red) line.

In addition, the resonance frequency for fixed drive power shifts with magnetic field—owing to the fact that the effective inductance of the nanowire depends on its critical current, which in turn depends on the magnetic field. At low temperature and low power, this resonant frequency shift can be as large as ~ 5 MHz (Fig. 16). When the drive power is large enough to give rise to a crater, this frequency shift has the same periodicity as that of $P_c$ (see Fig. 15) and is continuous. At lower input powers, however, the dependence of the resonance frequency on the magnetic field becomes discontinuous and multivalued (see Fig. 16). At higher powers or higher temperatures (for which the critical current is lower), the barrier to vortex entry is lower; hence one expects the system to return to periodic behavior, as is seen experimentally (see Fig. 16). We interpret the discontinuous and multivalued resonant frequency as a manifestation of the multivalued nature of the current-phase relationship of a long nanowire (i.e., $L/\xi > 4.4$) [41]: there are multiple possible metastable states differing in their values of the current circulating in the nanowires, corresponding to the presence of one or more virtual vortices trapped in the area between the wires; the resonance frequency of the resonator is thus shifted by an amount that is related to the number of vortices trapped between the wires, and the entry or exit of vortices corresponds to a jump in the resonance frequency.

The vortex-entry process in a resonator differs crucially from that in the d.c. geometries considered in Ref. [30], in the sense that the overall system is not maintained at a particular bias current: the minimal input current required to observe the resonance shift is far lower than the circulating current. The present a.c. approach therefore raises the prospect of *noninvasive* measurements of phase-slip and vortex dynamics in a two-wire device. It would thus be of considerable interest to perform analogous measurements on resonators containing thinner wires that are closer to the superconductor-insulator transition: we shall return to this idea in future work.

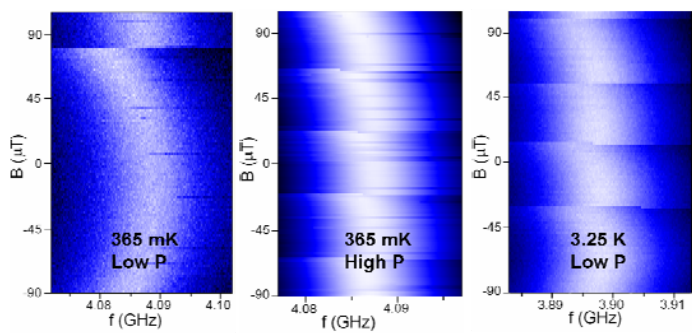

*Fig.* 16. Color map of transmission characteristics as a function of magnetic field and frequency for sample S5. The left graph is measured at 365 mK and an input power roughly 6 orders of magnitude less power than $P_c$. The middle graph is also measured at 365 mK but at only 3 dB less power than $P_c$. Here, periodicity returns, as the barrier for vortex jumping into/out of the loop formed by the two nanowiress is reduced. The right graph is measured at 3.25 K and using ~ 4 orders of magnitude less power than $P_c$. Again, periodicity returns, due to the reduction of the barrier for vortex jumping. Lighter (darker) color denotes higher (lower) transmission.

*Anomalous transmission in the frustrated state*

The scenario developed so far, viz., that the magnetic field affects the properties of the resonator by introducing a Meissner current and phase vorticies between the wires (where we understand a "phase vortex" to be a state in which the phase changes by $2\pi$ over a loop formed by the wires). The additional currents flowing through the wires add to the microwave-induced current thus reducing the power $P_c$ at which the crater appears. This leads one to expect that the crater should grow broader and deeper, monotonically, as the magnetic field is swept so as to decrease the critical power $P_c$. Although this expectation is borne out as regards the *width* of the crater, it is not borne out for the *depth*. Instead, for a narrow range of magnetic fields near the field corresponding to the lowest critical current (which we refer to as the fully "frustrated" state, i.e. states at which the loop cannot acquire the right number of phase vorticies to compensate the Meissner current; such a state occurs when the leads impose a phase difference of $2\pi n + \pi$ on the wires.), the crater becomes much flatter and shallower [see Fig. 17(a)], meaning that the supercurrent amplitude actually becomes large near this frustration point, which seems to go against expectations. Concomitantly, the satellite peaks in the transmitted power spectrum are broadened and decreased in height relative to the unfrustrated state, and a wide feature develops near the drive frequency [see Fig. 17(b) and (d)]. At higher temperatures, the satellite peaks disappear altogether at the frustration field.

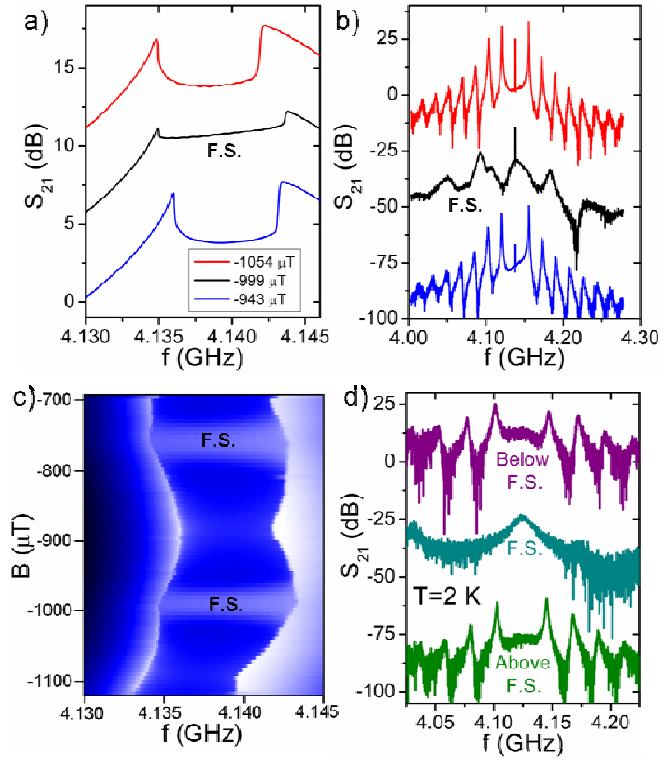

*Fig. 17.* (a) Transmission characteristics for sample S6, which is a two-wire device incorporating somewhat asymmetric nanowires separated by 1.65 μm and connecting a center conductor that is 10 μm in width, measured at 308 mK and at a power of -21 dB. In the frustrated state (denoted by F. S.), for which the critical power is at a minimum, an anomalous transmission effect is observed. For low input power and at zero magnetic field, $Q = 475$ for this sample. (b) The transmission power spectrum of the amplitude oscillations for drive frequency fixed at 4138 MHz and -20.8 dB. Near the frustrated state, the satellite peak spacing increases. (c) Color map of the transmission coefficient as a function of magnetic field and frequency, exhibiting the anomalously enhanced transmission

effect near the frustrated state (the corresponding regions are marked "F.S"). (d) The transmission power spectrum measured in the crater regime at high temperature (2 K) for drive frequency and power fixed at 4124 MHz and -39 dB. As one tunes the wires to the frustrated state (F. S.; middle curve), the satellite peaks vanish and are replaced by a broad central feature, thus indicating that the periodic superconductor-normal oscillations have been entirely replaced by stochastic dissipative events of a lesser strength. The terms "below" and "above" the F. S. indicate the magnetic field corresponding to the phase induced on the wires that is an integer multiple of $2\pi$, directly below and above the F. S., respectively.

This anomalous transmission effect is most evident at *higher* temperatures. Additionally, of the two samples that we measured, the more *asymmetric* sample (i.e., the one in which the wire critical currents are presumed less similar based on the differences in their physical appearance) exhibited a much more pronounced anomalous transmission feature (i.e. the rise of the bottom of the crater near the frustration point). Whereas, in the more symmetric sample, the anomalous transmission set in only at temperatures above $T \sim 2$ K (Fig. 18), in the asymmetric sample this effect persisted down to the lowest temperature at which we took data, i.e., $T = 308$ mK.

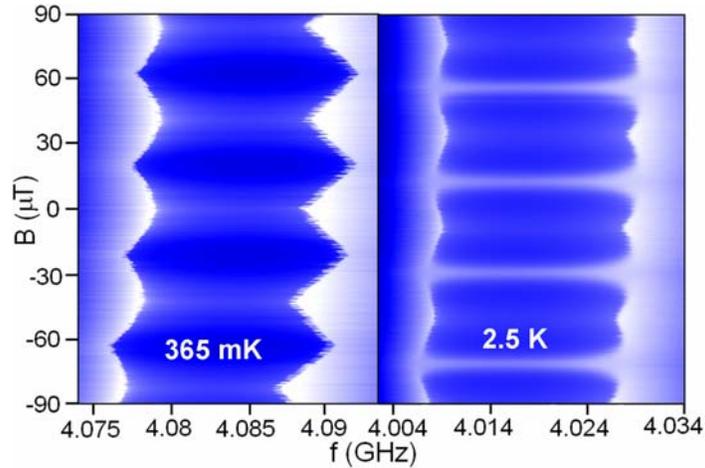

*Fig. 18.* Color maps of the transmission coefficient as a function of magnetic field and frequency, measured for an output power of -17 dB at the two indicated temperatures for the case of a device incorporating two symmetric nanowires (sample S5). The anomalous transmission effect only appears at higher temperatures.

*Qualitative explanation of the anomalous transmission*

The anomalous transmission effect seems to depend strongly on the asymmetry between the two wires. Therefore, a natural starting point for explaining it is to consider a resonator containing one wire that is *much* thicker than the other, i.e., the geometry shown in Fig. 19. This geometry is related to that of an rf SQUID that is capacitively coupled to the a.c. input. The analogy with an rf SQUID explains how a *plateau* in the transmission could arise: when the driving power exceeds a certain flux-dependent value, vortices are free to enter and leave the circuit via the weak link, thus dissipating energy. In contrast to the process discussed in Secs. 3 and 4 for the one-wire device, however, this dissipative process does not cause the *Q*-factor of the resonator to drop to zero when the weaker wire undergoes a dissipative process, as the two halves of the resonator remain connected by the stronger nanowire; instead, the total current in the resonator is expected to stay rather large in this regime [35]. This picture also accounts for the broad peak in frequency space near the drive frequency [see Fig. 17(d)], as the processes of vortex entry and exit are stochastic. Despite its idealization, therefore, this picture does qualitatively account for the high-temperature, asymmetric-wire data shown in Fig. 17(d).

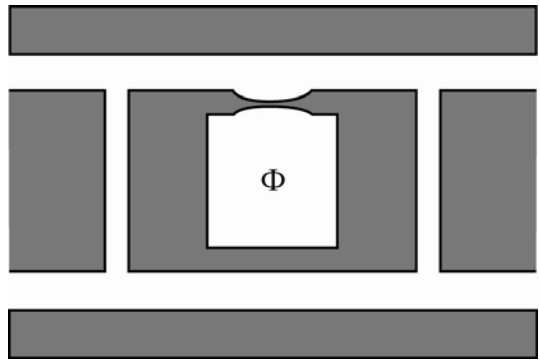

*Fig. 19.* Illustration of a two-wire sample in the "rf SQUID" limit of extreme asymmetry between the wires. The capacitive coupling between the resonator and its input is analogous to the coupling between an rf SQUID and its "tank" circuit [35].

The crucial difference between the scenario considered in the previous paragraph and the one considered in Secs. 3 and 4 is that in the former case the dissipative processes are concentrated in *one* of the wires. In the experimentally relevant regime, the disparities between the critical currents of the two wires are not as dramatic as in the limit considered in the previous paragraph; the "anomalous" dissipative state, in which only one wire becomes dissipative, must compete with the "normal" dissipative state, in which both wires become dissipative and the phenomenology described in Secs. 3 and 4 is realized. The threshold input power for the anomalous process is *lowest* in the frustrated state, whereas that for the normal process does not depend on the magnetic field; therefore, the anomalous process is most likely to occur in the frustrated state. In less asymmetrical situations, the weaker wire carries an appreciable fraction of the current in the circuit when superconducting; therefore, when the weaker wire switches to the normal state, one would expect the steady-state current in the circuit to drop to some fraction of its maximum achievable value. In particular, it is possible for the weaker wire *alone* to undergo amplitude oscillations (the stronger wire would always remain superconducting), in which case the current amplitudes in the circuit would oscillate between the low value, which equals the supercurrent in the stronger wire, and the high value corresponding to the addition of supercurrents from both wires.

These expectations are borne out by the data shown in Fig. 20. The fact that the satellite peaks spacing $\Delta f$ increases by a factor of ~ 3 in the frustrated state has a natural explanation in the scenario sketched in the present section: the current should take ~ 1/3 the time to grow from ~ $2I_c/3$ to $I_c$ as it does to grow from 0 to $I_c$. Therefore, the amplitude oscillation period should be reduced by a factor of ~ 1/3, and the frequency $\Delta f$ tripled, as the data shown in Fig. 20(b) indicates. We are also able to fit the shape of the anomalous crater by using the model presented in Sec. 4 but with a modified switching rule, in which the current is reduced to some fraction of $I_c$ (here ~ 2/3) rather than to zero. In the fits for the frustrated case, the diffusive parameter $t_{hold}$ is not included, so as to simplify the simulation. If $t_{hold}$ were included, it would serve to slightly increase the depth of the crater and would have forced (within the model) the supercurrent to drop to a slightly higher fraction of $I_c$, resulting in a slightly larger $\Delta f$. In the unfrustrated case the model predicts $\Delta f$ = 44 MHz, whereas the experimental value we have obtained is 50 MHz; in the frustrated state the model predicts $\Delta f$ = 12.4 MHz, whereas the experimental value is found to be 15 MHz. Therefore, the model has reasonable semi-quantitative agreement with the experiment.

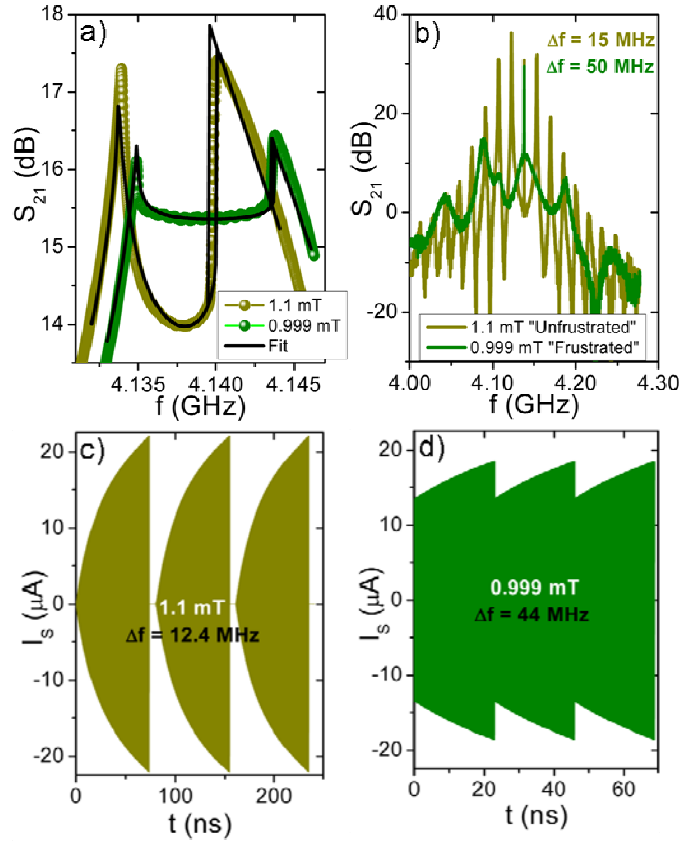

*Fig. 20.* (a) Transmission characteristics for a two-wire sample (S6) having noticeably asymmetric nanowires, measured at 308 mK and a power of -21 dB. In the frustrated state an anomalous transmission effect is observed. The model

described in Sec. 3 is used to fit the curve, but with a modified switching rule such that when the supercurrent hits $I_c$, it drops to a fraction of $I_c$, here, 0.64, because the stronger nanowire maintains superconductivity. The fits were calculated using the following fitting parameters: $C = 18.12$ pF, $R = 545$ Ω, $I_{cL} = 50.28$ μA; and for the unfrustrated (frustrated) state: $L = 0.07498$ (0.07493) nH; $I_c = 22.07$ (18.50) μA; $t_{hold} = 6.3$ (0) ns. Additionally, a small slope of 87 ndB/Hz was subtracted from each set of data to account for low temperature parasitic resonances as discussed in Sec. 4(c). (b) Transmission power spectrum of the amplitude oscillations for drive frequency fixed at 4138 MHz. In the frustrated state, the period of amplitude oscillations is reduced, thus increasing the satellite peak spacing $\Delta f$. (c) Model prediction for the supercurrent profile as a function of time at the unfrustrating magnetic field. (d) Model prediction for the supercurrent time evolution at the frustrating field; notice that the supercurrent does not drop to zero after it reaches $I_c$.

The effective inter-wire asymmetry—parameterized, e.g., by the ratio of the critical currents of the wires—is enhanced at temperatures that are high enough to be comparable to the critical temperature of the weaker of the two wires; thus, the anomalous transmission effect becomes more pronounced at higher temperatures. (The critical temperature is related to the critical current by the Bardeen formula, as discussed in Sec. 3.) A further, relatively minor, factor is that at high temperatures the thermal current noise in the resonator becomes comparable to the nanowire's critical current; thus, the deterministic amplitude-growth process delineated in Sec. 4 becomes swamped by the effects of thermal fluctuations. (Thus, one sees behavior that is superficially similar to anomalous transmission at temperatures around 4 K, even in the single-wire case. This effect cannot, however, account for the existence of anomalous transmission presented in this section because (i) at 300 mK in the asymmetric sample the thermal noise is much lower at this temperature and (ii) as the magnetic field is swept away from the frustrated state, the crater is observed to *increase* and not *decrease* in depth, which would be the wrong trend if the anomalous transmission effect were not present.

## 6. Conclusions and future directions

In the present work we have characterized a simple "many-body" circuit QED system, viz., a microwave stripline resonator interrupted by one or two nanowire bridges. We have identified two nonequilibrium steady states: one, which we have identified as an oscillatory steady state of the resonator-nanowire system, in which the nanowire periodically enters and leaves the superconducting state; and a second, stochastic steady state, which emerges in the two-wire device near what we have termed "frustrating" magnetic fields, and which we conjecture to be associated with vortex (or equivalently: phase slip) motion across the weaker of the two wires. We have presented evidence for the fact that the oscillatory steady state exists in a range of resonators containing quasi-one-dimensional elements, and is associated with the driving of the nanowire (or other quasi-one-dimensional element) being past its critical current. In addition, we have developed a simple phenomenological model that explains the salient features of the oscillatory steady state, and also captures some qualitative features of the stochastic steady state. Moreover, whilst accounting for the features of the oscillatory state, our model also enables us to extract information about the relaxation of heat pulses in nanowires; we find that, contrary to what one might expect, this relaxation does not slow down appreciably at temperatures far below $T_c$, but rather it saturates. We have also offered a qualitative picture of the "anomalous" stochastic state exhibited by two-wire devices, a feature that we hope to address in more detail in future work.

We believe that the primary avenue for future investigations of nanowires embedded in superconducting resonators should involve the study of nanowires that are much narrower than those measured to date. As discussed in the Introduction, such nanowires would have critical currents that are not much greater than the current due to a single photon in the resonator. Therefore, resonators containing them could be used both to investigate quantum phase-slips via a novel probe and to explore many-body circuit QED, in which a single photon is coupled to the elementary excitations of an *extended* quantum-mechanical system. Such devices would differ from the artificial-atom-based systems studied to date in a variety of ways; we briefly mention two. First, it has been predicted [42] that successive quantum phase-slip events in nanowires are coherent at low temperatures. As discussed in Refs. [43, 44], such coherence gives rise to an effective energy band structure for the states of the field representing charge transfer across the wire; this effective band structure is accompanied by interband "excitonic" transitions having frequencies in the microwave regime [43]. It is plausible that one could detect such excitonic transitions—which would provide strong evidence for the coherent quantum-mechanical character of phase slips—via their influence on cavity resonances, which would include, e.g., vacuum Rabi splitting [3]. Second, the physics of a single photon coupled to a quantum field (e.g., the superconducting phase fluctuations of the nanowire) would pave the way for realizations of quantum impurity-like models in which the *photon* acts as the impurity and the nanowire acts as a (one-dimensional) environment or bath. Quantum impurity models are believed to exhibit nonperturbative phenomena of considerable theoretical interest, such as the Kondo effect; moreover, the coupling of a low-dimensional system to a controllable "impurity" has been proposed as a method for probing the quantum mechanics of low-dimensional systems [45].

**Acknowledgments**

This material is based upon work supported by the U.S. Department of Energy, Division of Materials Sciences under Award No. DE-FG02-07ER46453, through the Frederick Seitz Materials Research Laboratory at the University of Illinois at Urbana-Champaign and through the NSF grant DMR 10-05645. S.G. was supported in party by the National Science Foundation under Grant No. NSF PHY05-51164 and NSF DMR10-05645.

---

[1] A. Wallraff, D. I. Schuster, A. Blais, L. Frunzio, R.-S. Huang, J. Majer, S. Kumar, S. M. Girvin and R. J. Schoelkopf. Nature **431**, 162 (2004); R. J. Schoelkopf and S. M. Girvin, Nature **451**, 664 (2008).

[2] By circuit QED we mean situations in which the excitations of the mesoscopic elements interact with the photons of the resonator, thus generating coupled nonequilibrium states of the composite system.

[3] See, e.g., S. Haroche and J.-M. Raimond, *Exploring the Quantum: Atoms, Cavities, and Photons* (Oxford University Press, 2006).

[4] V. Bouchiat, D. Vion, P. Joyez, D. Esteve, and M. H. Devoret, Phys. Scr. **T76**, 165 (1998).

[5] Y. Aharonov and L. Vaidman, arXiv:quant-ph/0105101v2 (2007).

[6] V. E. Manucharyan, E. Boaknin, M. Metcalfe, R. Vijay, I. Siddiqi, and M. Devoret, Phys. Rev. B **76**, 014524 (2007).

[7] M. Metcalfe, E. Boaknin, V. Manucharyan, R. Vijay, I. Siddiqi, C. Rigetti, L. Frunzio, R. J. Schoelkopf, and M. H. Devoret, Phys. Rev. B **76**, 174516 (2007).

[8] N. Bergeal, F. Schackert, M. Metcalfe, R. Vijay, V. E. Manucharyan, L. Frunzio, D. E. Prober, R. J. Schoelkopf, S. M. Girvin & M. H. Devoret, Nature **465**, 64 (2010).

[9] W. A. Little, Phys. Rev. **156**, 396 (1967).

[10] N. Giordano, Phys. Rev. Lett. **61**, 2137 (1988).

[11] C. N. Lau, N. Markovic, M. Bockrath, A. Bezryadin, and M. Tinkham, Phys. Rev. Lett. **87**, 217003 (2001).

[12] A. Bezryadin, C. N. Lau, and M. Tinkham, Nature **404**, 971 (2000).

[13] M. Sahu, M.-H. Bae, A. Rogachev, D. Pekker, T.-C. Wei, N. Shah, P. M. Goldbart, and A. Bezryadin, Nature Physics **5**, 503-508 (2009).

[14] D. Pekker, N. Shah, M. Sahu, A. Bezryadin, and P. M. Goldbart, Phys. Rev. B **80**, 214525 (2009).

[15] N. Shah, D. Pekker, and P. M. Goldbart, Phys. Rev. Lett. **101**, 207001 (2008)

[16] J. Ku, V. Manucharyan, and A. Bezryadin, Phys. Rev. B **82**, 134518 (2010).

[17] F. Brennecke, T. Donner, S. Ritter, T. Bourdel, M. Köhl, and T. Esslinger, Nature **450**, 268 (2007).

[18] P. Domokos and H. Ritsch, Phys. Rev. Lett. **89**, 253003 (2002); J. K. Asbóth, P. Domokos, H. Ritsch, and A. Vukics, Phys. Rev. A **72**, 053417 (2005).

[19] K. Baumann, C. Guerlin, F. Brennecke, and T. Esslinger, Nature **464**, 1301 (2010).

[20] S. Gopalakrishnan, B. L. Lev, and P. M. Goldbart, Nature Physics **5**, 845 (2009).

[21] I. B. Mekhov, C. Maschler, and H. Ritsch, Nature Physics **3**, 319 (2007).

[22] A. C. Anderson, R. C. Withers, S. A. Reible, R. W. Ralston, IEEE Trans. on Mag. **19**, 485 (1983).

[23] M. S. DiIorio, A. C. Anderson, B. –Y. Tsaur, Phys. Rev. B **38**, 7019 (1988).

[24] L. Frunzio, A, Wallraff, D. Schuster, J. Majer, R. Schoelkoph, IEEE Trans. on Appl. Superc. **15**, 860 (2005).

[25] E. Boaknin, V. Manucharyan, S. Fissette, M. Metcalfe, L. Frunzio, R. Vijay, I. Siddiqi, A. Wallraff, R. Schoelkopf, and M. H. Devoret, arXiv: 0702445 (2007).

[26] A. Bezryadin, J. Phys. C **20,** 043202 (2008).

[27] A. Bezryadin, A. Bollinger, D. Hopkins, M. Murphey, M. Remeika, and A. Rogachev, "Superconducting Nanowires Templated by Single Molecules," review article in *Dekker Encyclopedia of Nanoscience and Nanotechnology*, James A. Schwarz, Cristian I. Contescu, and Karol Putyera, eds. (Marcel Dekker, Inc. New York, 2004), 3761–3774.

[28] T. Aref and A. Bezryadin, arXiv:1006.5760v1 (2010).

[29] E. Segev, B. Abdo, O. Shtempluck, and E. Buks, J. Phys. C **19**, 096206 (2007).

[30] D. S. Hopkins, D. Pekker, P. M. Goldbart, and A. Bezryadin, Science **308**, 1762 (2005).

[31] D. S. Hopkins, D. Pekker, T.-C. Wei, P. M. Goldbart, and A. Bezryadin, Phys. Rev. B **76**, 220506 (2007).

[32] The quality factor is also observed to drop for higher order modes as it depends on $1/f_0$, where $f_0$ is the resonance frequency.

[33] J. Bardeen, Rev. Mod. Phys. **34,** 667 (1962).

[34] S. J. Hedges, M. J. Adams, and B. F. Nicholson, Electronics Letters **26**, 977 (1990).

[35] Michael Tinkham, *Introduction to superconductivity* (Dover, 2004).

[36] M. J. Lighthill, *An Introduction to Fourier Analysis and Generalised Functions* (Cambridge University Press, 1958).

[37] A. H. Nayfeh and D. T. Mook, *Nonlinear Oscillations* (Wiley-Interscience, New York, 1979).

[38] A. Lupascu, E. F. C. Driessen, L. Roschier, C. J. P. M. Harmans, and J. E. Mooij, Phys. Rev. Lett. **96**, 127003 (2006).

[39] E. Segev, O. Suchoi, O. Shtempluck, and E. Buks, Appl. Phys. Lett. **95**, 152509 (2009).

[40] D. Pekker, A. Bezryadin, D. S. Hopkins, and P. M. Goldbart, Phys. Rev. B **72**, 104517 (2005).

[41] K. K. Likharev, Rev. Mod. Phys. **51**, 101 (1979).

[42] H. P. Büchler, V. B. Geshkenbein, and G. Blatter, Phys. Rev. Lett. **92**, 067007 (2004).

[43] S. Khlebnikov, Phys. Rev. B **78**, 014512 (2008).

[44] J. E. Mooij and Yu. V. Nazarov, Nat. Phys. **2**, 169 (2006).

[45] A. Recati, P. O. Fedichev, W. Zwerger, J. von Delft, and P. Zoller, Phys. Rev. Lett. **94**, 040404 (2005).